\documentstyle[aps,12pt,epsf,manuscript]{revtex}
\newcommand{\bfsigma}{\mbox{\boldmath$\sigma$}}
\tightenlines
\begin{document}

\draft 

\title{Kaonic modes in hyperonic matter \\
and {\boldmath$p$}-wave kaon condensation}
\author{Takumi Muto\thanks{Email address; muto@pf.it-chiba.ac.jp}}
\address{Department of Physics, Chiba Institute of Technology, \\
        2-1-1 Shibazono, Narashino, Chiba 275-0023, Japan}

\date{\today}
\maketitle
\vspace{1.0cm}
      
\begin{abstract}
Kaon excitations (kaonic modes) are investigated in hyperonic 
matter, where hyperons ($\Lambda$, $\Sigma^-$, $\Xi^-$ ) are mixed in 
the ground state of neutron-star matter. $P$-wave kaon-baryon 
interactions as well as the $s$-wave interactions are taken into 
account within chiral effective Lagrangian, and  the nonrelativistic 
effective baryon-baryon interactions are incorporated. When the 
hyperon $\Lambda$ is more abundant than the proton at high baryon  
density, a proton-particle-$\Lambda$-hole mode, which has the $K^+$ 
quantum number, appears in addition to other particle-hole modes with 
the $K^-$ quantum number. It is shown that the system becomes 
unstable with respect to a spontaneous 
creation of a pair of the particle-hole modes with $K^+$ and 
$K^-$ quantum numbers, stemming from the $p$-wave kaon-baryon 
interaction.  The onset density of this $p$-wave kaon condensation 
may be lower than that of the $s$-wave $K^-$ condensation. 
\end{abstract}

\begin{description} 
{\footnotesize\item PACS: 05.30.Jp, 11.55.Fv, 13.75.Jz, 26.60.+c
\item Keywords:  kaon-baryon interaction, neutron stars, dispersion 
relations, kaon condensation}
\end{description}

\newpage
\section{Introduction}
\label{sec:intro}

Kaon condensation in high density matter has been investigated  
extensively from various points of 
view\cite{kn86,lbm95,t95,fmmt96,l96,pbpelk97}. 
Its existence makes the equation of state (EOS) of high density 
matter much softened,  which affects static properties of neutron 
stars\cite{fmmt96,tpl94,mfmt94,gs98} and also 
dynamical evolution of protoneutron stars\cite{ty98,p00}. A kaon 
condensate also leads to rapid cooling of neutron stars via enhanced 
neutrino emissions\cite{tpl94,bkpp88,t88,pb90,fmtt94}. 
The driving force for kaon condensation is the $s$-wave kaon-nucleon 
($K N$) interaction which consists of the scalar attraction 
simulated by the $KN$ sigma term $\Sigma_{KN}$ and the vector 
interaction corresponding to the Tomozawa Weinberg term.  In 
neutron-star matter, the lowest excitation energy $\omega_{\rm min}$ 
of the antikaon decreases with baryon number density $\rho_{\rm B}$ 
due to the $s$-wave $KN$ attraction. At a critical density,  
$\omega_{\rm min}$ becomes equal to the kaon chemical potential 
$\mu_K$, and the $K^-$ appears macroscopically as a Bose-Einstein 
condensate (BEC).\footnote{The kaon chemical potential $\mu_K$ is 
equal to the electron chemical potential $\mu_e$ in chemical 
equilibrium. We put $\mu_K=\mu_e=\mu$, and call $\mu$ the charge 
chemical potential throughout this paper.}
The critical density $\rho_{\rm B}^{\rm C}$ has been estimated as 
$\rho_{\rm B}^{\rm C} =3-4\rho_0$ with $\rho_0$ (=0.16 fm$^{-3}$)  
being the nuclear saturation 
density\cite{kn86,lbm95,t95,fmmt96,l96,pbpelk97}. 
Recently, some authors examined the possibility of kaon 
condensation in neutron stars by taking into account many-body 
effects such as the Pauli-blocking and the nucleon-nucleon 
correlations\cite{wrw97,chp00}. 

Studies of kaon condensation stimulated theoretical investigations  
of in-medium kaon properties with reference to the kaon-nucleon 
scattering\cite{k94,wkw96,l98,ro00}, 
kaonic atoms\cite{fgb94,fgm99,ho00,brn00}, 
and heavy-ion collisions\cite{llb97,kl96,slk99,smb97,cb99}. The 
momentum dependence of kaon optical potentials in a nuclear medium 
has also been discussed, which may serve as unified understanding of 
these phenomena\cite{sc98,ske00,trp00}. Although there is still a 
debate concerning the strength of the kaon optical potential, recent 
experimental results on the subthreshold $K^+K^-$ production in 
relativistic heavy-ion collisions and  proton-nucleus collisions 
suggest a substantial decrease in the antikaon effective 
mass\cite{b97,l99,ss99}. Based on the strongly attractive kaonic 
potential, a possibility of kaonic nuclei has been 
proposed\cite{k99,a00,fg00}. 

So far, kaon condensation has been considered mostly in neutron-star matter 
consisting of neutrons, protons and leptons in chemical 
equilibrium. However, hyperons ($\Lambda$, $\Xi^-$, 
$\Sigma^-$$\cdots$) may appear in neutron-star matter. Since the early 
suggestion of hyperon-mixing in neutron stars\cite{c59,as60}, 
the possible existence of hyperonic matter, where hyperons are mixed 
as well as neutrons, protons, and leptons in the ground state of 
neutron-star matter, has been discussed by several 
authors\cite{tc66,lr70,p71,g85,ekp95,sm96,s00,h98,bbs98,v00,bg97,blc99,y00,phz99,h00}. 

Recent development of hypernuclear experiments enables us to discuss 
the hyperon-mixing problem in highly dense matter in a realistic 
situation. It has been shown that hyperons appear at a baryon number 
density $\rho_{\rm B}=2\sim 3\rho_0$ based on the relativistic 
mean-field (RMF) models\cite{g85,ekp95,sm96,s00}, relativistic 
Hartree-Fock methods\cite{h98}, the nonrelativistic reaction matrix 
theory\cite{bbs98,v00}, the nonrelativistic effective baryon-baryon 
potential models\cite{bg97,blc99,y00}, etc.\cite{phz99,h00}
The hyperonic matter is also relevant to static and dynamic 
properties of neutron stars\cite{kj95,p99} and thermal evolution of 
neutron stars via rapid cooling\cite{ppl92}. 

The existence region of kaon condensation and hyperons may overlap 
each other, so that the competition or coexistence problem of these 
phases has to be clarified. Concerning this problem, it has been 
pointed out that a critical density of $K^-$ condensation, which may 
be realized from hyperonic matter, is pushed up to a higher baryon 
number density in the presence of the negatively charged hyperons in 
comparison with the neutron-star matter consisting of only neutrons, 
protons, and leptons\cite{ekp95,sm96}: 
The negative charge of the system is carried by the negatively 
charged hyperons in place of the electrons as the hyperon-mixing 
develops, making up with the positive charge and satisfying the 
charge neutrality. Accordingly, the number of the electrons $\rho_e$ 
and the charge chemical potential $\mu$ [=$(3\pi^2 \rho_e)^{1/3}$] get 
smaller with the increase in the baryon number density, which makes 
the onset condition for kaon condensation, $\omega_{\rm min}=\mu$, 
difficult to satisfy\footnote{The same situation is applied for the 
neutral hyperons, e.g., $\Lambda$: As the $\Lambda$ appears, the 
electron is absorbed by the process, $p+e^-\rightarrow \Lambda+\nu_e$ 
(see \ref{subsec:frac}). }. In Refs.~\cite{ekp95,sm96},  only the $s$-wave 
kaon-baryon interaction has been taken into account. 
However, there is the $p$-wave kaon-baryon interaction with Yukawa couplings, 
which may affect kaon dynamics in dense matter crucially as well as 
the $s$-wave interaction.  In this context, interrelation between kaons 
and hyperons has been discussed through the introduction of a ``kaesobar'' 
which is a linear combination of the $K^-$ and $\Sigma^-$-particle-neutron-hole 
states produced by the $p$-wave interaction\cite{llb97,blr98}.
On the other hand, 
the effects of the $p$-wave kaon-baryon interaction on kaon 
condensation have been considered in neutron-star matter where 
hyperons are not mixed in the ground state of dense 
matter\cite{m93,kvk95}. In Ref.~\cite{m93}, the kaon-baryon 
interaction has been taken into account within chiral effective 
Lagrangian. It has been shown that the $s$-wave $K^-$ condensation 
sets in at a lower density, and at a higher density, the $p$-wave 
$K^-$ condensation is realized accompanying hyperon excitation in the 
form of quasi-particles which are superimposed states of nucleons and 
hyperons. The $K^-$ mode has been shown to be relevant to the onset 
of $p$-wave $K^-$ condensation.\footnote{Throughout this paper, the 
$K^-$ and $K^+$ denote the kaonic modes which reduce to free kaons in 
vacuum ($\rho_{\rm B}\rightarrow 0$), and are distinguished from 
other particle-hole modes carrying the $K^\pm$ quantum numbers. }
In Ref.~\cite{kvk95}, pionic intermediate states, kaon fluctuations 
and residual interaction have been incorporated in addition to the 
$s$-wave and $p$-wave kaon-baryon interactions. It has been shown 
that the energy of a low-lying $\Lambda$-particle-proton-hole branch 
becomes negative beyond  some density, leading to a first-order phase 
transition to proton matter accompanying kaon condensation. 
 
In this paper, we consider kaon dynamics and mechanisms of kaon 
condensation in {\it hyperonic matter} by taking into account the 
$p$-wave kaon-baryon interaction. We discuss in-medium properties of 
kaons by obtaining the kaon dispersion relations in 
hyperonic matter. Kaon-baryon interactions for both the $s$-wave and 
$p$-wave type are incorporated within the effective chiral 
Lagrangian, while other baryon-baryon interactions are supplemented 
by the use of the nonrelativistic effective interactions. We show 
that a low-lying proton-particle-$\Lambda$-hole mode with the $K^+$ 
quantum number develops at high densities. This mode, together with 
the other particle-hole modes carrying the $K^-$ quantum number,  
leads to a new mechanism of kaon condensation stemming from the 
$p$-wave kaon-baryon interaction.\footnote{
Part of this work has been briefly reported in Ref.~\cite{m00}.}

The paper is organized as follows. Section~\ref{sec:form} gives the 
formulation to obtain the kaonic modes in hyperonic matter. In 
Sec.~\ref{sec:result}, numerical results and discussion are given. Summary and 
concluding remarks are devoted in Sec.~\ref{sec:summary}. In the 
Appendix, the expression for the potential energy density which is 
used in the formulation is presented. 

\section{Formulation}
\label{sec:form}
\subsection{Kaon-baryon interaction}
\label{subsec:kbint}

We start with the effective chiral SU(3)$_L \times$ SU(3)$_R$   
Lagrangian used by Kaplan and Nelson\cite{kn86} for the kaon-baryon 
interaction. \footnote{
We use the units in which $\hbar$=$c$=$k_B$=1 throughout this paper.}
\begin{eqnarray}
{\cal L}&=&\frac{1}{4}f^2 \ {\rm Tr} 
\partial^\mu\Sigma^\dagger\partial_\mu\Sigma 
+\frac{1}{2}f^2\Lambda_{\chi{\rm SB}}({\rm Tr}M(\Sigma-1)+{\rm h.c.}) 
\cr
&+&{\rm Tr}\overline{\Psi}(i{\not\partial}-M_{\rm B})\Psi+{\rm 
Tr}\overline{\Psi}i\gamma^\mu\lbrack V_\mu, \Psi\rbrack
+D{\rm Tr}\overline{\Psi}\gamma^\mu\gamma^5\{A_\mu, \Psi\}
+F{\rm Tr}\overline{\Psi}\gamma^\mu\gamma^5\lbrack A_\mu, \Psi\rbrack 
\cr
&+&a_1{\rm Tr}\overline{\Psi}(\xi M^\dagger\xi+{\rm h.c.})\Psi 
+ a_2{\rm Tr}\overline{\Psi}\Psi(\xi M^\dagger\xi+{\rm h.c.})+
a_3({\rm Tr}M\Sigma +{\rm h.c.}){\rm Tr}\overline{\Psi}\Psi \ , 
\label{eq:lag}
\end{eqnarray}
where $f$($\sim f_\pi$=93 MeV) is the meson decay constant, $\Sigma$ 
is the nonlinear meson field, $\Sigma\equiv e^{2i\Pi/f}$, in terms of 
the octet meson $\Pi$, which is represented as 
\begin{eqnarray}
\Pi=\pi_aT_a=\frac{1}{\sqrt{2}}\left(
\begin{array}{ccc}
\pi^0/\sqrt{2}+\eta/\sqrt{6} & \pi^+ & K^+ \\
\pi^- & -\pi^0/\sqrt{2}+\eta/\sqrt{6} & K^0 \\
K^- & \overline{K}^0 & -\sqrt{\frac{2}{3}}\eta \\
\end{array}\right)
\label{eq:meson}
\end{eqnarray}
with the Nambu-Goldstone bosons $\pi_a$ and the SU(3) generators 
$T_a$ ($a=1\sim 8$).  $\Lambda_{\chi{\rm SB}}$ is the chiral symmetry 
breaking scale, $\sim$ 1 GeV, $M$ the mass matrix which is defined as 
$M\equiv {\rm diag}(m_u, m_d, m_s)$ with the quark masses $m_i$, 
and $M_{\rm B}$ is the baryon mass generated as a consequence of 
spontaneous chiral symmetry breakdown.  The baryon fields are given 
as the spin 1/2 octet $\Psi$, 
\begin{eqnarray}
\Psi=\left(
\begin{array}{ccc}
\Sigma^0/\sqrt{2}+\Lambda/\sqrt{6}  & \Sigma^+ &  p  \\
\Sigma^- & -\Sigma^0/\sqrt{2}+\Lambda/\sqrt{6} &  n    \\
\Xi^- & \Xi^0 & -2\Lambda/\sqrt{6}       \\
\end{array}\right) \ . 
\label{eq:psi}
\end{eqnarray}
 Baryons couple with mesons through the mesonic vector current 
$V_\mu$, which is defined by $V_\mu\equiv 
1/2(\xi^\dagger\partial_\mu\xi+\xi\partial_\mu\xi^\dagger)$ with 
$\xi\equiv \Sigma^{1/2}$, and the  axial vector current $A_\mu$, 
defined by $A_\mu\equiv 
i/2(\xi^\dagger\partial_\mu\xi-\xi\partial_\mu\xi^\dagger)$. In the 
nonrelativistic limit, 
the fourth term including $V_\mu$ in Eq.~(\ref{eq:lag}) reduces to 
the $s$-wave 
vector interaction, and the fifth and sixth terms to the $p$-wave 
interaction. The last three terms in Eq.~(\ref{eq:lag}) represent the 
$s$-wave meson-baryon scalar interaction, which explicitly breaks 
down chiral symmetry. 
Throughout this paper, we take the same set of parameters as those in 
Ref.~\cite{kn86}: The axial vector coupling constants $D$ and $F$ are 
determined from the weak decays of the baryon octet states and are 
chosen as $D$=0.81 and $F$=0.44.  
 The quark masses are taken to be $m_u$=6 MeV, $m_d$=12 MeV and 
$m_s$=240 MeV. With these values for $m_i$, the parameters $a_1$ and 
$a_2$ are fixed so as to give the empirical octet baryon mass 
splittings and given as $a_1$=$-$0.28, $a_2$=0.56. 
 
 The parameter $a_3$ is related to the ``$\pi N$ and $KN$ sigma 
terms" which simulate the $s$-wave meson-baryon scalar interactions 
through the relations, $\Sigma_{\pi N}=-(a_1+2a_3)(m_u+m_d)$, 
$\Sigma_{Kp}=-(a_1+a_2+2a_3)(m_u+m_s)$, 
and $\Sigma_{Kn}=-(a_2+2a_3)(m_u+m_s)$. 
 Considering that there is some ambiguity about these values, we  
take  $a_3$ to be $a_3=-0.9$ and $-1.1$, which yields 
$\Sigma_{\pi N}$=37 MeV, $\Sigma_{Kp}$=374 MeV, 
$\Sigma_{Kn}$=305 MeV for $a_3=-0.9$, and $\Sigma_{\pi N}$=45 MeV, 
$\Sigma_{Kp}$=472 MeV, $\Sigma_{Kn}$=403 MeV for $a_3=-1.1$. 

 In this paper, we mainly concentrate on the kaon dispersion 
relations in hyperonic matter and onset mechanisms of kaon 
condensation. 
Expecting that we can also discuss the EOS 
of a fully-developed kaon-condensed phase beyond the onset 
density\cite{m01} within the same framework, we give here a 
formulation to elucidate both issues in a unified way: First, the 
energy density of the kaon-condensed phase, ${\cal E}_{\rm eff}$,  is 
obtained, and second, the kaon inverse propagator $D_K^{-1}(\omega, 
{\bf k}; \rho_B)$ is obtained  from the  expansion of the energy 
density expression with respect to the classical kaon field $K^{\rm 
cl}$ as \break
$\displaystyle {\cal E}_{\rm eff}(|K^{\rm cl}|)
={\cal E}_{\rm eff}(|K^{\rm cl}|=0)-D_K^{-1}(\omega =\mu, {\bf k}; 
\rho_B) |K^{\rm cl}|^2+O(|K^{\rm cl}|^4)$.\footnote{
Similar methods have been utilized in Refs.~\cite{fmmt96,m93}. }

The kaon inverse propagator $D_K^{-1}$ includes the self energy which 
is related to the forward $KN$ scattering amplitudes. 
It is to be noted that there are different ways of off-shell 
extrapolation from the $KN$ scattering amplitudes in obtaining the 
kaon self energy: One is derived 
from chiral perturbation theory and the other is based on low energy 
theorems of current algebra and PCAC (partial conservation of 
axial-vector current). These approaches lead to different off-shell 
properties of the kaon excitation 
energy\cite{ynm93,ljm94,lbm95,fmmt96}. The relations between these 
approaches have been discussed\cite{ymk94,tw95}. 
Based on the former way with the chiral effective Lagrangian 
Eq.~(\ref{eq:lag}), the $s$-wave $KN$ scattering amplitudes near 
threshold cannot be reproduced empirically without higher order terms 
in chiral expansion and a pole contribution from the 
$\Lambda(1405)$ (abbreviated to $\Lambda^\ast$)\cite{ljm94,lbm95,wkw96}. 
These corrections enter into 
the kaon self energy.  Nevertheless, it has been shown 
that part of the higher order terms for $K^-$, the range term 
[$\propto \omega(K^-)^2$] , 
becomes small in high-density matter, since the $K^-$ excitation 
energy $ \omega(K^-)$ decreases with density owing to the $s$-wave 
$KN$ attractive interaction. Furthermore, the pole contribution from 
the $\Lambda^\ast$ becomes negligible since the $K^-N$ threshold lies 
far below the pole of the $\Lambda^\ast$ at high density due to the decrease 
in $ \omega(K^-)$. \footnote{
On the other hand, the role of the $\Lambda^\ast$ in a nuclear medium has been 
discussed on the assumption that it is a $K^-p$ bound state\cite{k94,wkw96,l98,ro00}: 
In Ref.~\cite{k94}, it has been shown that the mass of the $\Lambda^\ast$ 
is shifted upwards due to the Pauli blocking of intermediate states, 
and that the $K^-$ feels an attractive potential by scattering through 
an intermediate $\Lambda^\ast$ state. In subsequent works, the mass of the 
$\Lambda^\ast$ has been shown to be left as a free space value 
due to the net effects of the Pauli-blocking and the attraction which comes 
from the modification of the $K^-$ in the intermediate $K^-N$ states, 
while the decay width of the $\Lambda^\ast$ increases with an increase 
in the baryon number density\cite{l98,ro00}. }
Hence these corrections become irrelevant to the $s$-wave $K^-$ 
excitation energy in high-density 
matter\cite{lbm95,fmmt96,wkw96,ljm94}, while the $K^+$ excitation 
energy is largely changed due to the repulsive effects from the 
higher-order corrections. 

When hyperons are incorporated, extra terms 
next to leading order, coming from the $p$-wave kaon-baryon 
interaction, contribute to the self energy. These terms are 
typically  proportional to $(k^\mu)^2=\omega^2-{\bf k}^2$, and are 
responsible for reproducing the low-energy data on pion- and 
photon-induced  kaon production and elastic and inelastic $K^- p$ 
scatterings\cite{rkww00,lk00}. 
In hyperonic matter, 
several kaonic modes of particle-hole excitations, $\Lambda p^{-1}$, 
$\Sigma^- n^{-1}$, $\Xi^- \Lambda^{-1}$ , appear in addition 
to the $K^\pm$ due to the $p$-wave interaction, as shown in 
Sec.~\ref{sec:result}\footnote{The superscript `$-1$' denotes a hole state. }, 
and the excitation energies for these modes 
are small, being of order of the $s$-wave $K^-$ energy. Thus we 
expect that the higher order terms coming from the $p$-wave 
interaction have minor effects on the kaonic modes as far as the kaon momentum 
is not very large, i.e., $|{\bf k}|=O(\omega)$ except for the $K^+$. 
Hence, throughout this paper, we are based on the 
chiral effective Lagrangian Eq.~(\ref{eq:lag}), and don't take into 
account higher order terms in chiral expansion with respect to 
$M/\Lambda_{\chi SB}$ and $\partial/\Lambda_{\chi SB}$. 

There are also other $p$-wave subthreshold 
resonances such as $\Sigma(1385)$ (abbreviated to $\Sigma^\ast$) , 
which may appear in the dispersion relations of the kaonic modes. 
However, the excitation energies of the $\Sigma^\ast N^{-1}$ modes are 
of the order $\sim$450 MeV so that 
their branches lie far above the other particle-hole branches, 
$p\Lambda^{-1}$, $\Sigma^-n^{-1}$, $\Xi^-\Lambda^{-1}$ considered in this paper.
Furthermore, their coupling strengths $g_{\Sigma^\ast N}$ to the $K^- N$ are 
not so large as compared with $g_{\Lambda p}$, $g_{\Sigma^- n}$, 
and $g_{\Xi^-\Lambda}$. Hence the inclusion of these resonances would not 
change the results quantitatively. 
Quantitative evaluation including the higher order terms in chiral perturbation 
and effects of the subthreshold resonances remains to be a future investigation. 

\subsection{Effective energy density}
\label{subsec:energy}

For simplicity, we take only the  $\Lambda$, $\Sigma^-$, and $\Xi^-$ 
for hyperons in addition to the proton ($p$), neutron ($n$) for 
nucleons and the ultrarelativistic electron for leptons. The other 
hyperons, $\Sigma^0$, $\Sigma^+$ and $\Xi^0$, are supposed to be 
irrelevant because they appear in higher densities due to their heavy 
baryon masses and  due to the fact that the electron chemical 
potential does not assist to satisfy the threshold condition 
in contrast to the case of the negatively charged hyperons. 

We  consider charged kaon condensation in chemically-equilibrated 
hyperonic matter, and neglect other mesons in the octet meson field 
$\Sigma$. 
The classical kaon field is then assumed to be a plane wave type:
\begin{equation}
K^{\pm, {\rm cl}}({\bf r}, t)=\frac{f}{\sqrt{2}}\theta e^{\pm i(\mu_K 
t-{\bf k}\cdot {\bf r})} \ , 
\label{eq:classical}
\end{equation}
where $\theta$ is a chiral angle which means an amplitude of a  
condensate, $\mu_K$ the kaon chemical potential, and ${\bf k}$ is the 
kaon momentum. 

The effective Hamiltonian ${\cal H}_{\rm eff}$ is derived with the 
introduction of the charge chemical potential $\mu$(=$\mu_K$=$\mu_e$) 
by a charge neutrality condition, and is separated into baryon, meson 
and lepton parts: 
\begin{equation}
{\cal H}_{\rm eff}={\cal H}_{\rm eff}^{\rm B}+{\cal H}_{\rm eff}^{\rm 
M} +{\cal H}_{\rm eff}^e \ , 
\label{eq:eff}
\end{equation}
where $\displaystyle {\cal H}_{\rm eff}^{\rm B}={\cal H}^{\rm 
B}+\mu\rho_Q^{\rm B}$, ${\cal H}_{\rm eff}^{\rm M}={\cal H}^{\rm 
M}+\mu\rho_Q^{\rm M}$, and ${\cal H}_{\rm eff}^e ={\cal 
H}^e+\mu\rho_Q^e$ with the charge density operators $\rho_Q^i$ 
($i={\rm B}, {\rm M}, e$). 

The axial vector coupling terms in the baryonic part ${\cal H}_{\rm 
eff}^{\rm B}$ have space-time dependent factors $\exp(\pm ip_K\cdot 
x)$ with a four momentum $p_K^\mu$ [=($\mu_K, {\bf k})$]. These 
factors are eliminated by transformation of the baryon bases 
$\psi^T=(p, \Lambda, \Xi^-, n, \Sigma^- )$ as $\psi^T\rightarrow 
\tilde\psi^T=(e^{-ip_K\cdot x}p, \Lambda, e^{ip_K\cdot x}\Xi^-, 
e^{-ip_K\cdot x/2}n, e^{ip_K\cdot x/2}\Sigma^-)$.  By this 
transformation, the momentum of each baryon is shifted by 
${\cal V}_3 {\bf k}$, where ${\cal V}_3$ is the  $V$-spin defined by 
$\displaystyle 
{\cal V}_3=\frac{1}{2}\Big(I_3+\frac{3}{2}Y\Big)$ with the third 
component 
of the isospin $I_3$ and the hypercharge $Y$. After making the 
Foldy-Wouthuysen-Tani transformation for ${\cal H}_{\rm eff}^{\rm B}$ 
and expanding it up to $O(1/M_N)$ with $M_N$ the nucleon mass, one 
obtains the baryonic part of the Hamiltonian in a nonrelativistic 
form as ${\widetilde{\cal H}}_{\rm eff}^{\rm B}={\tilde\psi}^\dagger 
H_{\rm eff}^{\rm B} {\tilde\psi}$ with 
\begin{eqnarray}
H_{\rm eff}^{\rm B}=\left(
\begin{array}{ccccc}
H_{pp} & H_{p\Lambda} & H_{p\Xi^-} & 0 & 0 \\
H_{\Lambda p} & H_{\Lambda \Lambda} & H_{\Lambda\Xi^-} &  0 & 0 \\
H_{\Xi^- p} & H_{\Xi^-\Lambda} & H_{\Xi^-\Xi^-} & 0 & 0 \\
0 & 0 & 0 & H_{nn} & H_{n\Sigma^-}  \\
0 & 0 & 0 & H_{\Sigma^- n} & H_{\Sigma^-\Sigma^-} \\
\end{array}\right) \ , 
\label{eq:bhm}
\end{eqnarray}
where the nonvanishing matrix elements are given by 
\begin{eqnarray}
& &H_{pp}=\frac{1}{2M_N}\Big\{({\bf p}-{\bf 
k}\cos\theta)^2+\Big(\frac{1}{2}g_{\Lambda 
p}\mu\sin\theta\Big)^2\Big\}+\mu\cos\theta-\Sigma_{Kp}(1-\cos\theta) 
\cr
& &H_{p\Lambda}=\frac{i}{2}g_{\Lambda p}\sin\theta\Big\{{\bf 
k}-\frac{\mu}{2M_N}(2{\bf p}-{\bf k}\cos\theta)\Big\}
\cdot\bfsigma=H_{\Lambda p}^\ast \cr
& & H_{p\Xi^-}=\frac{1}{8M_N} g_{\Lambda p} 
g_{\Xi^-\Lambda}(\mu\sin\theta)^2 =H_{\Xi^- p} \cr
& &H_{\Lambda\Lambda}=\frac{1}{2M_N}\Big\lbrack{\bf p}^2+
\Big\{\Big(\frac{g_{\Lambda p}}{2}\Big)^2
+\Big(\frac{g_{\Xi^- 
\Lambda}}{2}\Big)^2\Big\}(\mu\sin\theta)^2\Big\rbrack+
 \delta M_{\Lambda N}-\Sigma_{K\Lambda}(1-\cos\theta) \cr
& &H_{\Lambda\Xi^-}=-\frac{i}{2}g_{\Xi^-\Lambda}\sin\theta
\Big\{{\bf k}-\frac{\mu}{2M_N}(2{\bf p}+{\bf k}\cos\theta)\Big\}\cdot 
\bfsigma=H_{\Xi^-\Lambda}^\ast\cr
& &H_{\Xi^-\Xi^-}=\frac{1}{2M_N}\Big\{({\bf p}+{\bf 
k}\cos\theta)^2+\Big(\frac{1}{2}
g_{\Xi^-\Lambda}\mu\sin\theta\Big)^2\Big\}+\delta M_{\Xi^- 
N}-\mu\cos\theta-\Sigma_{K\Xi^-}(1-\cos\theta) \cr
& &H_{nn}=\frac{1}{2M_N}\Big\{\Big({\bf p}-\frac{1}{2}{\bf 
k}\cos\theta\Big)^2+\Big(\frac{1}{2}
g_{\Sigma^- 
n}\mu\sin\theta\Big)^2\Big\}-\mu\sin^2\frac{\theta}{2}-\Sigma_{Kn}(1-\cos\theta) 
\cr
& &H_{n\Sigma^-}=-\frac{i}{2}g_{\Sigma^- n}\sin\theta
\Big({\bf k}-\frac{\mu}{M_N}{\bf p}\Big)\cdot\bfsigma=H_{\Sigma^- 
n}^\ast \cr
& &H_{\rm \Sigma^- \Sigma^-}=\frac{1}{2M_N}\Big\{\Big({\bf 
p}+\frac{1}{2}{\bf k}\cos\theta\Big)^2
+\Big(\frac{1}{2}g_{\Sigma^- n}\mu\sin\theta\Big)^2\Big\}  
+ \delta M_{\Sigma^- 
N}-\mu\cos^2\frac{\theta}{2}-\Sigma_{K\Sigma^-}(1-\cos\theta) \ . 
\label{eq:bh}
\end{eqnarray}
In Eq.~(\ref{eq:bh}),  $\Sigma_{KY}$ ($Y=\Lambda, \Sigma^-, \Xi^-$) 
are the ``kaon-hyperon sigma terms" given by 
$\displaystyle\Sigma_{K\Lambda}\equiv 
-\Big(\frac{5}{6}a_1+\frac{5}{6}a_2+2a_3\Big)
(m_{\rm u}+m_{\rm s})$, 
$\Sigma_{K\Sigma^-}=\Sigma_{Kn}$, and $\Sigma_{K\Xi^-}=\Sigma_{Kp}$. 
$\delta M_{Y N}$ ($Y=\Lambda, \Sigma^-, \Xi^-$) are the 
hyperon-nucleon mass difference. The $p$-wave axial vector coupling 
strengths are defined by $g_{\Lambda p}=(D+3F)/\sqrt{6}$ (=0.87), 
$g_{\Sigma^- n}=D-F$ (=0.37), and $g_{\Xi^- 
\Lambda}=(-D+3F)/\sqrt{6}$ (=0.21). For simplicity, the nucleon mass 
is taken to be the free neutron mass, $M_N=M_P=M_n$=939.57 MeV, and 
the mass differences $\delta M_{YN}$ are taken from the free baryon 
masses:  
$M_\Lambda$=1115.7 MeV, $M_{\Sigma^-}$=1197.4 MeV, and 
$M_{\Xi^-}$=1321.3 MeV. The $p$-wave kaon-baryon Yukawa couplings are 
brought about through the terms proportional to ${\bf 
k}\cdot\bfsigma$ with $\bfsigma$ the spin matrix in the off-diagonal 
matrix elements in Eq.~(\ref{eq:bh}). 
There appear terms of order $O(1/M_N)$ coming from meson-baryon 
recoils, ${\bf p}\cdot{\bf k}/M_N$, ${\bf k}^2/M_N$, 
$\mu {\bf p}\cdot\bfsigma/M_N$, $\mu^2\sin^2\theta/M_N$, etc. in 
Eq.~(\ref{eq:bh}). 
The magnitude of the baryon momentum $|{\bf p}| $ is typically the 
Fermi momentum $p_F$, which is in the range 
$p_F\lesssim 4m_\pi$ with $m_\pi$ being the pion mass. Thus 
these terms are considered to be small 
as compared with the kinetic energy ${\bf p}^2/(2M_N)$ of each baryon 
so long as $\mu=O(m_\pi)$ and $|{\bf k}|\lesssim 3 m_\pi$. 
Numerically, these terms are not necessarily small. Nevertheless, 
we would like to elucidate mechanisms of kaon condensation within  
the basic $p$-wave kaon-baryon interaction in hyperonic matter, at 
the cost of detailed corrections to the effective Hamiltonian. Hence, 
for simplicity, we neglect the terms of order $O(1/M_N)$ except for the kinetic energy term. 
 
 After diagonalization of $H_{\rm eff}^{\rm B}$,  the baryonic 
eigenstates are represented as ``quasiparticles" which consist of 
superposition of the baryons, e.g., $|\tilde 
p\rangle=\alpha|p\rangle+\beta |\Lambda\rangle +\gamma 
|\Xi^-\rangle$, 
  $|\tilde n\rangle=\delta | n\rangle+\epsilon |\Sigma^-\rangle$, 
etc. , where $\alpha$, $\beta$, $\gamma$, $\delta$, $\epsilon$ are 
coefficients.\footnote{
  In general, the $\Sigma^0$ state is also superposed to the 
quasiparticle states  $|\tilde p\rangle$, $|\tilde \Lambda\rangle$, 
and $|\tilde \Xi^-\rangle$ through the $p$-wave couplings, but we 
simply neglect the effects of the $\Sigma^0$ on these quasiparticle 
states. } 
The baryon  contribution to the effective energy density, ${\cal 
E}_{\rm eff}^{\rm B}$, is obtained from occupation of the 
quasiparticles over each Fermi sea\cite{m01}. 
\begin{equation}
{\cal E}_{\rm eff}^{\rm B}=\sum_i \sum_{\stackrel{|{\bf p}| \leq 
|{\bf p}_F(i)|}{s=\pm1/2}}E_s^{(i)} ({\bf p}) \ , 
\label{eq:be}
\end{equation}
where ${\bf p}_F(i)$ ($i={\tilde p}, {\tilde\Lambda}, {\tilde\Xi^-}, 
{\tilde n}, {\tilde \Sigma^-}$) are the Fermi momenta, and the 
subscript `$s$' stands for the spin states for the quasiparticles. 

The mesonic contribution to the effective energy density is given by 
the substitution of the classical kaon field (\ref{eq:classical}) 
into ${\cal H}_{\rm eff}^{\rm M}$ as 
\begin{equation}
\epsilon_{\rm eff}^{\rm M}=-\frac{1}{2}f^2(\mu^2-{\bf 
k}^2)\sin^2\theta+f^2m_K^2(1-\cos\theta) \ , 
\label{eq:me}
\end{equation}
where $m_K\equiv [\Lambda_{\chi {\rm SB}}(m_u+m_s)]^{1/2}$, which is 
identified with the free kaon mass, and is replaced by the 
experimental value, 493.7 MeV. The leptonic contribution  in 
Eq.~(\ref{eq:eff}) reduces to the effective energy density 
\begin{equation}
\epsilon_{\rm eff}^e
=\frac{\mu^4}{4\pi^2}-\mu\frac{\mu^3}{3\pi^2}
=-\frac{\mu^4}{12\pi^2}
\label{eq:ee}
\end{equation}
for the ultra-relativistic electrons. Thus one obtains the total 
effective energy density as ${\cal E}_{\rm eff}={\cal E}_{\rm 
eff}^{\rm B}+{\cal E}_{\rm eff}^{\rm M}+{\cal E}_{\rm eff}^e$. 

\subsection{Potential contribution}
\label{subsec:potential}

Here we introduce baryon-baryon interactions beyond the framework of 
chiral symmetry. The potential energy density ${\cal E}_{\rm pot}$ 
produced from baryon-baryon interactions is crucial to obtaining the 
EOS and matter composition of the ground state for not only the  
normal (noncondensed) phase but also the condensed phase. 
We suppose that ${\cal E}_{\rm pot}$ depends on the extent of kaon 
condensation only implicitly through the change of each number 
density $\rho_i$ ($i=p, \Lambda, \Xi^-, n, \Sigma^-$) due to 
condensation. 
Then the form of ${\cal E}_{\rm pot}$ is assumed to be ${\cal E}_{\rm 
pot}(\rho_p, \rho_\Lambda, \rho_{\Xi^-}, \rho_n, \rho_{\Sigma^-})$. 
Following a procedure for incorporating nuclear interactions in case 
of pion condensation\cite{ab74,mo88}, 
we define the potential for the baryon $V_i$ ($i$=$p$, $\Lambda$, 
$\Xi^-$, $n$, $\Sigma^-$) in hyperonic matter as  
\begin{equation}
 V_i=\partial {\cal E}_{\rm pot}/\partial \rho_i \ , 
 \label{eq:pot}
 \end{equation}
which corresponds to a potential contribution to each baryon chemical 
potential, i.e., estimated at the Fermi momentum of each baryon. Thus 
momentum dependence of the potential is neglected, which is 
reasonable for the $\Sigma^-$ and $\Lambda$ potentials  because of 
their weak momentum dependence, as suggested from recent microscopic 
calculations\cite{bbs98,v00}. It should be reminded,  however, that 
it is not a good approximation for the nuclear potentials where the 
potential depths at zero momentum are deeper than the value at the 
Fermi momentum. 

The potential terms $V_i$ are added to each diagonal matrix element 
of the baryonic part of the effective Hamiltonian 
$H_{\rm eff}^{\rm B}$ [ Eq.~(\ref{eq:bh}) ] as 
$\displaystyle H_{ii}\rightarrow H'_{ii}=H_{ii}+V_i$ .  Then the 
baryonic part of the modified 
effective energy density ${{\cal E}_{\rm eff}'}^{\rm B}$ is given by
\begin{equation}
{{\cal E}_{\rm eff}'}^{\rm B}=\sum_i \sum_{\stackrel{|{\bf p}| \leq 
|{\bf p}_F(i)|}{s=\pm1/2}}{E_s'}^{(i)} ({\bf p})+{\cal E}_{\rm 
pot}-\sum_{i=p, \Lambda, \Xi^-, n, \Sigma^-} \rho_i V_i\ , 
\label{eq:totaleb}
\end{equation}
where ${E_s'}^{(i)}({\bf p})$ are eigenvalues diagonalized with 
inclusion of the $V_i$ in the diagonal parts of $H_{\rm eff}^B$. The 
last term on the r. h. s. in Eq.~(\ref{eq:totaleb}) is introduced in 
order to subtract the double counting of the baryon interaction 
energies in the first sum 
over the quasiparticle energies ${E_s'}^{(i)}({\bf p})$. 

For a practical use of the potential energy density ${\cal E}_{\rm 
pot}$ in  hyperonic matter, we adopt the nonrelativistic expression 
by Balberg and Gal\cite{bg97}, which includes hyperon-hyperon 
interactions as well as hyperon-nucleon ones, and higher order terms 
in $\rho_i$  simulating the many-body effects. 
The expression for the ${\cal E}_{\rm pot}$ is given in Appendix A. 
We take the exponents $\delta$ and $\gamma$ in the density-dependent 
terms in Eq.~(\ref{eq:epot}) to be $\delta=\gamma=5/3$, which gives 
the moderate stiffness of the EOS among the three cases in 
Ref.~\cite{bg97}. 
Some of the parameters in ${\cal E}_{\rm pot}$ are refitted so as to 
be consistent with recent empirical data on nuclear and hypernuclear 
experiments and the saturation properties of symmetric 
nuclear matter:  We take the saturation density  $\rho_0$=0.16 
fm$^{-3}$, the binding energy 16 MeV, and the incompressibility 210 
MeV at $\rho_0$ in symmetric nuclear matter, from which $a_{NN}$ and 
$c_{NN}$ are fixed for the isoscalar  terms in the nucleon-nucleon 
($NN$) part of the potential energy [ Eq.~(\ref{eq:epot}) ]. From the 
symmetry energy $\sim$ 30 MeV at $\rho_B=\rho_0$, one obtains 
$b_{NN}$ for the isospin-dependent term for the $NN$ part.\footnote{
It is to be noted that the potential contribution $V_{\rm 
sym}(\rho_B)$ to the symmetry energy is read from the isovector term 
for the $NN$ interactions in 
Eq.~(\ref{eq:epot}) as $V_{\rm sym}(\rho_B)=b_{NN}\rho_B/2$,  which 
mimics the density dependence of the $V_{\rm sym}(\rho_B)$   in the 
RMF models. In normal neutron-star matter, this 
linear density dependence leads to a large proton fraction  
$\rho_p/\rho_B\gtrsim 0.1$ at high density and the large electron 
chemical potential $\mu_e$ through the relation 
$\mu_e=(3\pi^2\rho_e)^{1/3}$ and the charge neutrality 
$\rho_e=\rho_p$. The large proton fraction concerns with a 
possibility of the direct Urca process in neutron 
stars\cite{b81,lpph91}, 
and the large electron chemical potential assists the onset of the 
negatively charged hyperons, e.g., the $\Sigma^-$,  through the 
chemical equilibrium condition, 
$\mu_n+\mu_e=\mu_{\Sigma^-}$\cite{y00}. 
However, there is an ambiguity about the density dependence of the 
symmetry energy. E.g., the nonrelativistic potential models show    
its moderate increase in density in comparison with the RMF 
case\cite{fmtt94,lpph91}. }
For the $\Lambda N$ part, $a_{\Lambda N}$ and $c_{\Lambda N}$ are 
taken to be the same as in Ref.~\cite{bg97}. The depth of the 
$\Lambda$ potential in nuclear matter is then equal to the  empirical 
value, i.e., $V_{\Lambda}$ ( $\rho_p=\rho_n=\rho_0/2$, 
$\rho_\Lambda=\rho_{\Xi^-}=\rho_{\Sigma^-}=0$)=$a_{\Lambda 
N}\rho_0+c_{\Lambda N}\rho_0^\gamma$=$-$27 MeV\cite{mdg98}. 
For the $\Xi^- N$ part, $a_{\Xi N}$ and $c_{\Xi N}$ are related with 
each other by the use of  the depth of the $\Xi^-$ potential in 
nuclear matter, which is deduced from the recent ($K^-, K^+$) 
experimental data as $-14- -20$ MeV\cite{f98,k00}. Here we put  
$V_{\Xi^-}$ ( $\rho_p=\rho_n=\rho_0/2$, 
$\rho_\Lambda=\rho_{\Xi^-}=\rho_{\Sigma^-}=0$)=
$a_{\Xi N}\rho_0+c_{\Xi N}\rho_0^\gamma$=$-$16 MeV. 
Further, the crossover density $\rho_{\rm co}$, 
where $V_{\Xi^-}(\rho_B=\rho_{\rm co})$=0, is taken to be equal to 
that for the $\Lambda$\cite{bg97}. 
As for the $V_{\Sigma^-}$,   recent ($K^-$, $\pi^\pm$) experiment at 
BNL\cite{b99} suggests a strong isospin dependence in the 
$\Sigma$-nucleus potential. It has been shown that the experimental 
data is compatible with the analysis by the Nijmegen model F for the 
baryon-baryon interaction, which gives a repulsive isoscalar part and 
a large positive isospin dependent term (Lane potential) for the 
$\Sigma$-nucleus potential\cite{d99}. The analysis of $\Sigma^-$ atom 
data also shows the repulsive $\Sigma^-$ potential\cite{mfgj95}. 
Theoretically, a repulsive $\Sigma$ potential in nuclear matter has 
been obtained from the $G$ matrix calculation based on the SU(6) 
quark model baryon-baryon interaction\cite{kf00}. Referring to these 
results, 
we first take the potential depth of the $\Sigma^-$ in nuclear matter 
to be {\it repulsive} as follows:  In Ref.~\cite{d99}, the $\Sigma^-$ 
potential has been parametrized as 
$\displaystyle 
V_{\Sigma^-}(k_{\Sigma})=V_0(k_{\Sigma})-\frac{1}{2}V_1(k_{\Sigma}) 
\cdot\frac{2Z-A}{A}$, where $k_{\Sigma^-}$ is the $\Sigma^-$ 
momentum, and $A$, $Z$ are the mass number and  the atomic number, 
respectively. The calculated values 
in Ref.~\cite{d99} for  $V_0$ and $V_1$ at $k_\Sigma$=0 in nuclear 
matter based on the Nijmegen model F 
are identified with the corresponding terms in our model such that 
$V_{\Sigma^-}(\rho_p=\rho_n=\rho_0/2, 
\rho_\Lambda=\rho_{\Xi^-}=\rho_{\Sigma^-}=0)=a_{\Sigma 
N}\rho_0+c_{\Sigma N}\rho_0^\gamma$=$V_0$=23.5 MeV, and 
$b_{\Sigma N}\rho_0=V_1/2$=40.2 MeV. We call this parametrization for 
the repulsive $\Sigma^-$ potential Case I. 
For the second case, the parameters $a_{\Sigma N}$, 
$b_{\Sigma N}$, and $c_{\Sigma N}$ are taken to be the same as in 
Ref.~\cite{bg97}. The depth of the $V_{\Sigma^-}$ is then equal to 
that of the $\Lambda$ as an extreme case for the attractive 
$\Sigma^-$ potential, i.e.,  $a_{\Sigma N}\rho_0+c_{\Sigma 
N}\rho_0^\gamma$=$-$27 MeV. We call this parametrization for the 
attractive $\Sigma^-$ potential Case II. 

The remaining parameters in the potential energy density ${\cal 
E}_{\rm pot}$, relevant to the hyperon-hyperon interactions,  
are taken to be the same as those in Ref.~\cite{bg97}. The numerical 
values for the parameters are summarized in Tables~\ref{tab:para} and 
\ref{tab:paras} in Appendix A. 

In general, there are additional off-diagonal  matrix elements in 
${\cal H}_{\rm eff}^{\rm B}$ in the presence of condensation.  These 
 off-diagonal matrix elements include the densities 
$\bar\rho_{p\Lambda}\equiv i\langle p^\dagger\bfsigma\cdot{\hat {\bf 
k}} \Lambda\rangle$, $\bar\rho_{\Lambda\Xi^-}\equiv i\langle 
\Lambda^\dagger\bfsigma\cdot{\hat {\bf k}} \Xi^-\rangle$, 
$\bar\rho_{n\Sigma^-}\equiv i\langle n^\dagger\bfsigma\cdot{\hat {\bf 
k}} \Sigma^-\rangle$, where $\langle\cdots \rangle$ means the ground 
state expectation value with ${\hat{\bf k}}\equiv{\bf k}/|{\bf k}|$. 
The strengths in these off-diagonal matrix elements are approximately 
related to the Landau-Migdal parameters, which simulate the 
short-range correlations between baryons, as is the case of pion 
condensation\cite{ab74,mo88}. However, the values of the 
Landau-Migdal parameters 
for the particle-holes including hyperons are hardly known both 
theoretically and experimentally. Hence, in this paper, we don't take 
into account these extra off-diagonal matrix elements, by putting 
emphasis on a brief discussion about onset mechanisms of the $p$-wave 
kaon condensation within a simple model. 

\subsection{Kaon propagation in hyperonic matter}
\label{subsec:self}
 
\begin{figure}[t]
\centerline{
\epsfxsize=0.8\textwidth\epsffile{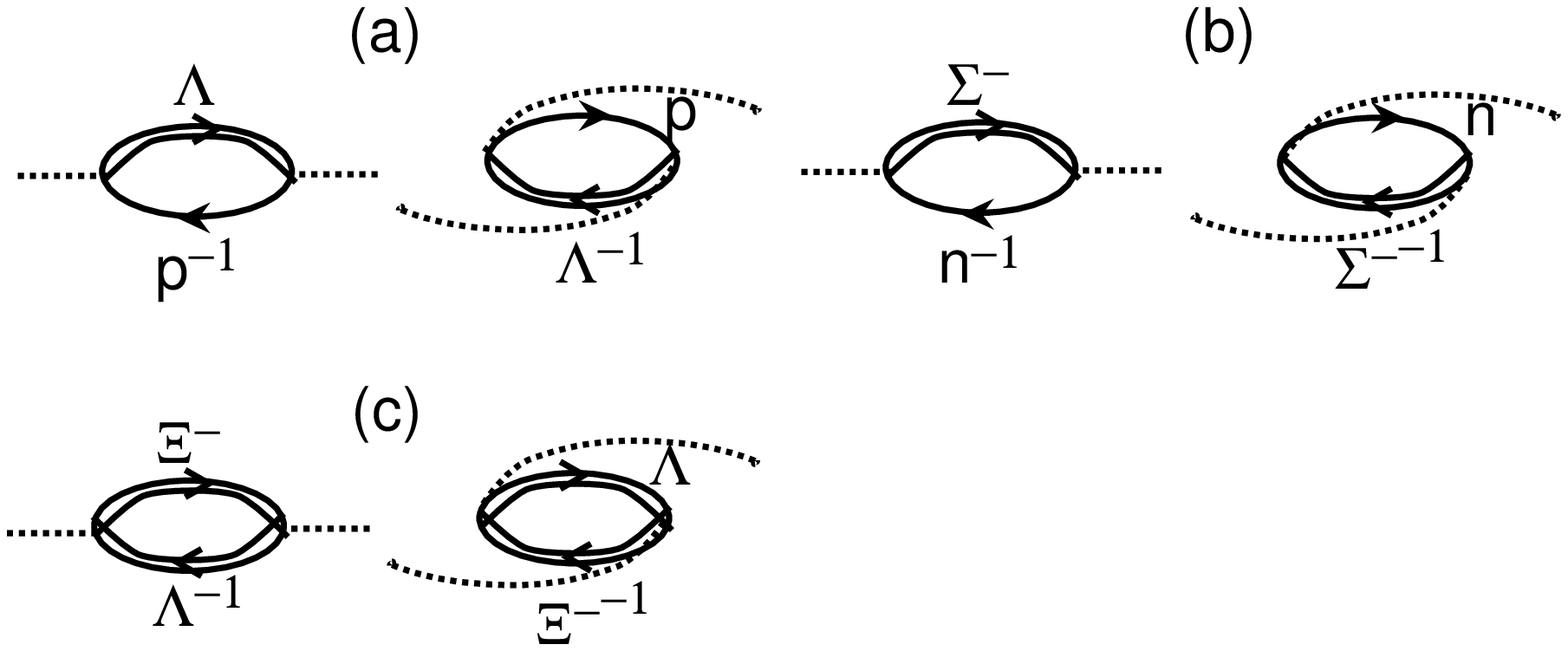}}
\caption{Pole contributions to the $K^-$ self energy from the 
$p$-wave kaon-baryon interactions:  (a) 
$\Lambda$-particle-proton-hole and proton-particle-$\Lambda$-hole 
states, (b) $\Sigma^-$-particle-neutron-hole 
and neutron-particle-$\Sigma^-$-hole states, and (c)
$\Xi^-$-particle-$\Lambda$-hole and 
$\Lambda$-particle-$\Xi^-$-hole states. }
\label{fig:pole}
\end{figure}

After expanding the total effective energy density ${\cal E}'_{\rm 
eff}$ ($\equiv{{\cal E}_{\rm eff}'}^{\rm B}+{\cal E}_{\rm eff}^{\rm 
M}+{\cal E}_{\rm eff}^e$) with respect to $\theta$ around $\theta=0$ 
as $\displaystyle {\cal E}'_{\rm eff}={\cal E}'_{\rm 
eff}(\theta=0)-\frac{f^2}{2}D_K^{-1}(\mu, {\bf k}; 
\rho_B)\theta^2+O(\theta^4)$, one obtains the kaon inverse 
propagator: 
\begin{equation}
D_K^{-1}(\omega, {\bf k};\rho_B)=
\omega ^2-{\bf k}^2-m_K^2 
-\Pi_{K}(\omega,{\bf k};\rho_B) 
\label{eq:dkinv}
\end{equation}
with the kaon self energy $\Pi_{K}(\omega,{\bf k};\rho_B)=
\Pi_{K}^s(\omega,{\bf k};\rho_B)+\Pi_{K}^p (\omega,{\bf k};\rho_B)$, 
where
\begin{mathletters}\label{eq:self}
\begin{eqnarray}
\Pi_{K}^s (\omega,{\bf k};\rho_B) &=&
-\frac{1}{f^2}\sum_i\rho_i\Sigma_{Ki} 
-\frac{1}{f^2}\Big(\rho_p+\frac{1}{2}\rho_n-\frac{1}{2}\rho_{\Sigma^-}-\rho_{\Xi^-}\Big)\omega  
\label{eq:selfs} \\
\Pi_{K}^p (\omega,{\bf k};\rho_B) &=&
-\frac{1}{2f^2}\Bigg\lbrack \frac{(\rho_p-\rho_\Lambda)(g_{\Lambda 
p}{\bf k})^2}{\delta M_{\Lambda p}-\omega +V_\Lambda-V_p}+
 \frac{(\rho_n-\rho_{\Sigma^-})(g_{\Sigma^- n}{\bf k})^2}{\delta 
M_{\Sigma^- n}-\omega +V_{\Sigma^-}-V_n} \cr
&+&\frac{(\rho_\Lambda-\rho_{\Xi^-})(g_{\Xi^-\Lambda}{\bf 
k})^2}{\delta M_{\Xi^-\Lambda}-\omega 
+V_{\Xi^-}-V_\Lambda}\Bigg\rbrack \ . \label{eq:selfp}
\end{eqnarray}
\end{mathletters}
The first term on the r.h.s. of Eq.~(\ref{eq:selfs}) gives the 
$s$-wave scalar attraction with $\rho_i$ the number densities for 
baryons $i$ ($i=p, \Lambda, \Xi^-, n, \Sigma^-$).   The second term 
on the r.h.s. of Eq.~(\ref{eq:selfs}) gives the $s$-wave vector 
interaction, where the coefficients in front of 
the number densities come from the $V$-spin (${\cal V}_3$). This term 
gives attraction for proton and neutron, while repulsion for 
$\Sigma^-$ and $\Xi^-$. 
The $p$-wave part $\Pi_K^p$ [Eq.~(\ref{eq:selfp})] consists of the 
pole contributions from the $p$-wave kaon-baryon interaction.  The 
diagrams corresponding to 
each term in Eq.~(\ref{eq:selfp}) are depicted in 
Fig.~\ref{fig:pole}.

The excitation energies for kaonic modes are obtained as zero points 
of the kaon inverse propagator, 
$D_K^{-1}(\omega, {\bf k};\rho_B)$, 
which depends on the composition of the ground state of noncondensed 
hyperonic matter, i.e., the number densities for the particles, 
$\rho_i$ ($i=p, \Lambda, \Xi^-, n, \Sigma^-, e^-$). The particle 
number densities are determined from 
charge neutrality condition, 
$\rho_p=\rho_{\Xi^-}+\rho_{\Sigma^-}+\rho_e $, 
baryon number conservation, 
$\rho_p+\rho_\Lambda+\rho_{\Xi^-}+\rho_n+\rho_{\Sigma^-}=\rho_B$, and 
chemical equilibrium conditions between $p$, $\Lambda$, $\Xi^-$, $n$, 
$\Sigma^-$ and $e^-$, 
\begin{mathletters}\label{eq:chemical}
\begin{eqnarray}
\mu_n&=&\mu_p+\mu_e \quad\quad \textrm{for}\quad n\rightleftharpoons 
pe^-{\bar\nu}_e \ ,  \label{eq:chemical1} \\
\mu_\Lambda&=&\mu_p+\mu_e \quad\quad\textrm{for}\quad 
pe^-\rightleftharpoons \Lambda\nu_e \ , \label{eq:chemical2}\\
\mu_{\Xi^-}&=&\mu_\Lambda+\mu_e \quad\quad\textrm{for}\quad \Lambda 
e^-\rightleftharpoons \Xi^-\nu_e \ , \label{eq:chemical3}\\
\mu_{\Sigma^-}&=&\mu_n+\mu_e \quad\quad\textrm{for}\quad 
ne^-\rightleftharpoons \Sigma^-\nu_e \ , \label{eq:chemical4}
\end{eqnarray}
\end{mathletters}
where the chemical potentials for baryons $\mu_i$ are given by 
\begin{equation}
 \mu_i=(3\pi^2\rho_i)^{2/3}/(2M_N)+
\delta M_{i N}+V_i  \quad(\delta M_{i N}=0 \ \textrm{for} \  i=p,n) . 
\label{eq:chempot} 
\end{equation}

\section{Numerical results and discussion}
\label{sec:result}
\subsection{$K^-$ optical potential}
\label{subsec:vopt}

We take a preliminary view of the in-medium kaon properties within 
our model by estimating a $K^-$ optical potential 
$V_{\rm opt}({\bf k}; \rho_B)$ at $\rho_B=\rho_0$ in symmetric 
nuclear matter. The $K^-$ optical potential $V_{\rm opt}({\bf k}; 
\rho_B)$ is defined in terms of the kaon self energy 
$\Pi_K$ [Eq. (\ref{eq:self}) ] as 
\begin{equation}
V_{\rm opt}({\bf k}; \rho_B)=\Pi_{K^-}
\Big(\omega({\bf k}, \rho_B),{\bf k}; \rho_B\Big)
/2 \omega({\bf k}, \rho_B) \ , 
\end{equation}
where $ \omega({\bf k}, \rho_B)$ is the $K^-$ excitation energy 
obtained from the dispersion equation, $D_K^{-1}(\omega,{\bf k}; 
\rho_B)=0$ at given $\rho_B$ and ${\bf k}$. 

\begin{figure}[t]
\centerline{
\epsfxsize=0.5\textwidth
\epsffile{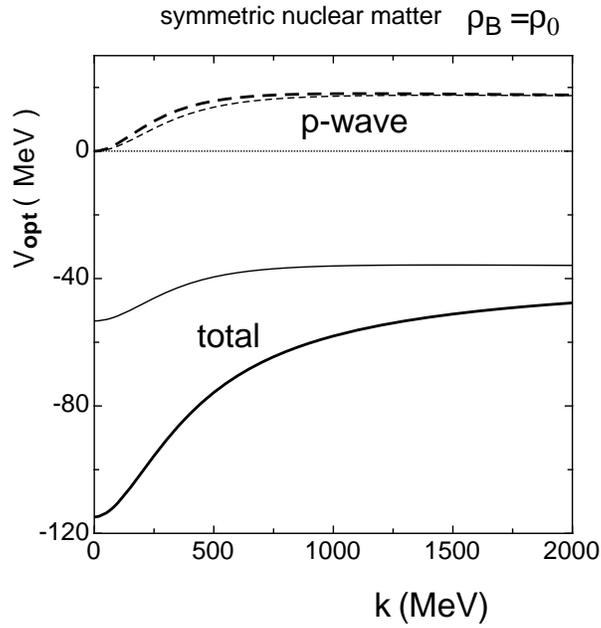}}
\caption{The $K^-$ optical potential at $\rho_B=\rho_0$ in symmetric 
nuclear matter.}
\label{fig:vopt}
\end{figure}

 In Fig. \ref{fig:vopt}, we show $V_{\rm opt}({\bf k}; \rho_0)$ as a 
function of the kaon momentum $|{\bf k}|$. The solid lines represent 
the total values coming from both the $s$ and $p$-wave interactions 
[see Eq. (\ref{eq:self}) ], and the dashed lines represent the 
contribution from the $p$-wave interaction. The bold lines are for 
$a_3=-0.9$ ($\Sigma_{Kn}$=305 MeV), and the thin lines for 
$a_3=-0.28$ ($\Sigma_{Kn}$=0). Since the numerical result depends 
little on the choice of the $\Sigma^-$ potential (Case I or Case 
II),  only  Case I is shown in the figure. The total attractive 
potential energy decreases monotonically with ${\bf k}$. 
The potential depth at zero momentum is  $\sim -$120 MeV for 
$a_3=-0.9$ and $\sim -55$ MeV for $a_3=-0.28$. 
It is to be noted that the total potential energy is attractive even 
if the $s$-wave scalar interaction is almost absent (thin solid line)
 due to the existence of the $s$-wave $K^- N$ vector 
attraction (the Tomozawa-Weinberg term). Since the excitation energy 
for $K^-$ is larger than those at the 
$\Sigma^- $ and $\Lambda$ poles, i.e., 
$\delta M_{\Sigma^- n}- \omega +V_{\Sigma^-}-V_n <0$ and 
$\delta M_{\Lambda p}- \omega +V_\Lambda-V_p <0$, 
the $p$-wave part of the optical potential is repulsive for $K^-$ at 
this density, as seen from Eq.~(\ref{eq:self}), and has a minor 
contribution in  comparison with the $s$-wave attraction. 
The $K^-$-nucleus potential has been obtained phenomenologically 
from kaonic atom data in several works. In Ref.~\cite{fgb94}, a strongly 
attractive potential inside a nucleus, 
Re $V_{\rm opt}=-200\pm 20$ MeV, has been obtained with a nonlinear density 
dependent term.  On the other hand, the energy shifts and widths have been 
well reproduced in Ref.~\cite{ho00} with much reduced attraction $\sim -45$ 
MeV in a local density approximation. 

Several authors have elaborated the momentum-dependent $K^-$ optical potential 
in nuclear matter. With a  coupled channel 
approach based on chiral models\cite{ske00,ro00}, or a $G$-matrix 
method with the J\" ulich $\bar K N$ interaction\cite{trp00} taking 
into account the in-medium modification of the kaon, they have 
obtained more moderate momentum dependence for Re $V_{\rm opt}$ than ours 
over the relevant momentum region. On the other hand, our result for $a_3=-0.9$ 
is similar to the result by a dispersion relation approach in Ref.~\cite{sc98}. 
At present,  there is a controversy about the magnitude of the $K^-$ potential 
at finite momentum as well as at zero momentum. 

\subsection{Particle fractions in hyperonic matter}
\label{subsec:frac}

\begin{figure}[t]
\noindent\begin{minipage}[l]{0.50\textwidth}
\noindent\centerline{
\epsfxsize=\textwidth\epsffile{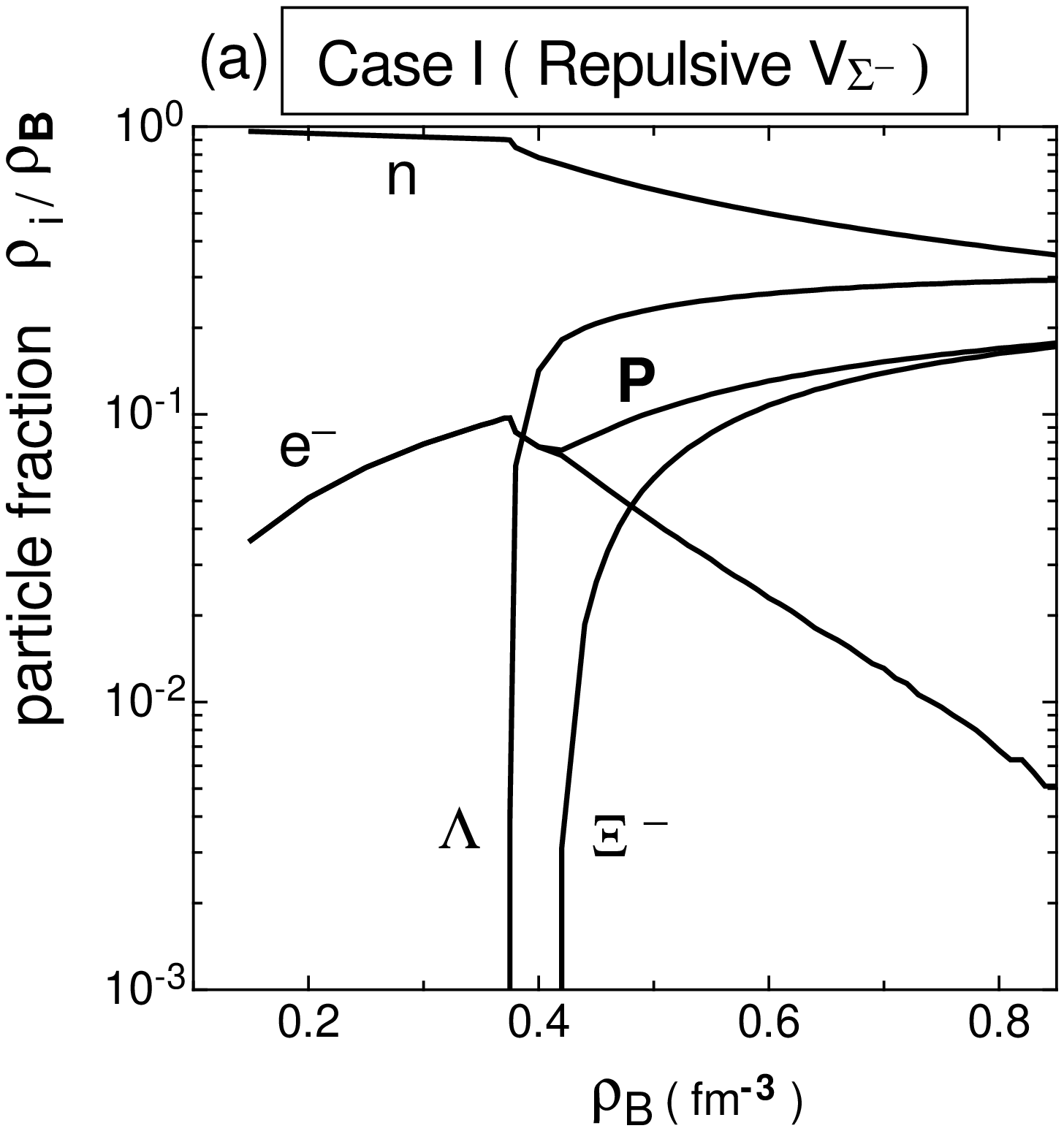}}
\end{minipage}~
\begin{minipage}[r]{0.50\textwidth}
\noindent\centerline{
\epsfxsize=\textwidth\epsffile{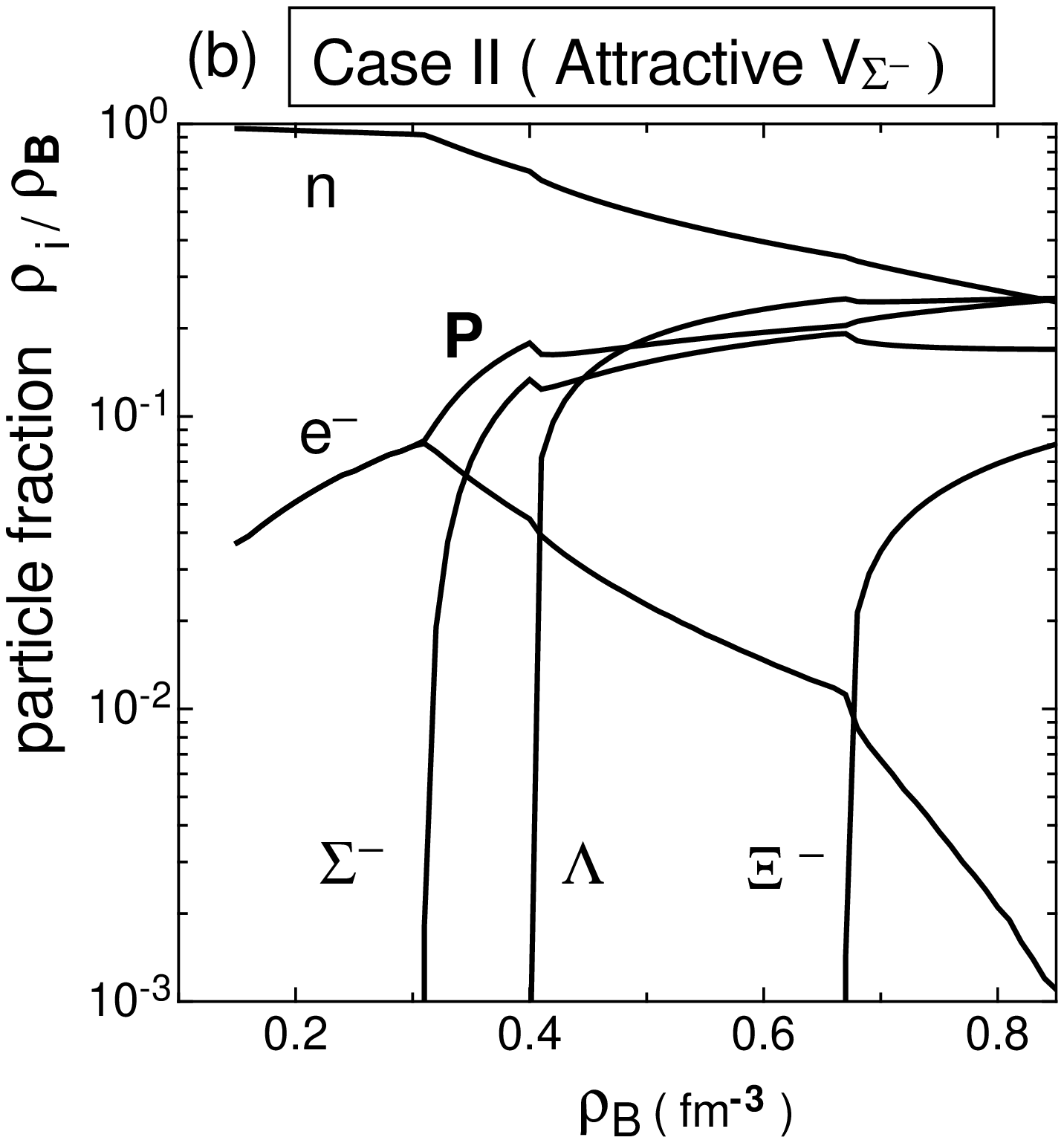}}
\end{minipage}
\caption{Particle fractions $\rho_i/\rho_{\rm B}$ as functions of the 
baryon number density 
$\rho_{\rm B}$ for (a) Case I (the repulsive $V_{\Sigma^-}$)  and 
(b) Case II (the attractive $V_{\Sigma^-}$).  }
\label{fig:frac}
\end{figure}

Before considering the behaviors of kaonic modes, we make a survey of 
the ground state properties of  
hyperonic matter by obtaining the matter composition $\rho_i$ 
($i=p,n,\Lambda,\Sigma^-,\Xi^-, e^-$),  which enters into the kaon 
self energy Eq.~(\ref{eq:self}), being responsible for kaon dynamics. 
The particle fractions $\rho_i/\rho_{\rm B}$ as functions 
of the baryon number density $\rho_B$ are shown in 
Fig.~\ref{fig:frac}~(a) for Case I (the repulsive $V_{\Sigma^-}$) and 
Fig.~\ref{fig:frac}~(b) for Case II (the attractive $V_{\Sigma^-}$). 

In Case I, the $\Lambda$ appears at $\rho_B\sim $ 0.37 fm$^{-3}$, 
and its fraction rapidly increases with density, exceeding the proton 
fraction ($\rho_\Lambda > \rho_p$) at $\rho_{\rm B}\sim$ 0.40 
fm$^{-3}$. Soon after the appearance of $\Lambda$, the $\Xi^-$ are 
mixed at $\rho_{\rm B}\sim$ 0.42 fm$^{-3}$, and it increases with 
density. On the other hand, the electron fraction rapidly decreases 
after the appearance of the $\Lambda$  and $\Xi^-$. The $\Sigma^-$ 
does not appear over the relevant densities because of its repulsive 
potential. 

In Case II, the $\Sigma^-$ first appears at $\rho_{\rm B}\sim$ 0.31 
fm$^{-3}$, and the $\Lambda$ appears at $\rho_{\rm B}\sim 0.40$ 
fm$^{-3}$ after the onset of $\Sigma^-$. Both fractions rapidly 
increase with density. In particular, the 
fraction of $\Lambda$ exceeds the proton fraction at $\rho_{\rm 
B}\sim$ 0.47 fm$^{-3}$. The electron fraction decreases due to the 
appearance of the negatively charged hyperon $\Sigma^-$, which is 
qualitatively the same feature as for Case I.  The electron chemical 
potential $\mu_e$ [$=(3\pi^2\rho_e)^{1/3}$] also becomes small with 
increase in density, which is unfavorable to matching the chemical 
equilibrium condition, $\mu_{\Xi^-}=\mu_\Lambda+\mu_e$. Thus the existence 
region of the $\Xi^-$ is pushed up to very high densities as compared 
with that in Case I. 

These results for matter composition in Case I and II qualitatively 
reproduce the results of Figs.~4 and  3 in Ref.~\cite{bg97}, 
respectively. 

\subsection{Kaonic modes in hyperonic matter}
\label{subsec:kaon}

\subsubsection{Case I (the repulsive $V_{\Sigma^-}$)}
\label{subsubsec:disprvsm}

\noindent\underline{(i) The weaker $s$-wave scalar interaction 
($a_3=-0.9$) }
\vspace{0.3cm}

\begin{figure}[h]
\noindent\begin{minipage}[l]{0.50\textwidth}
\centerline{
\epsfxsize=\textwidth
\epsffile{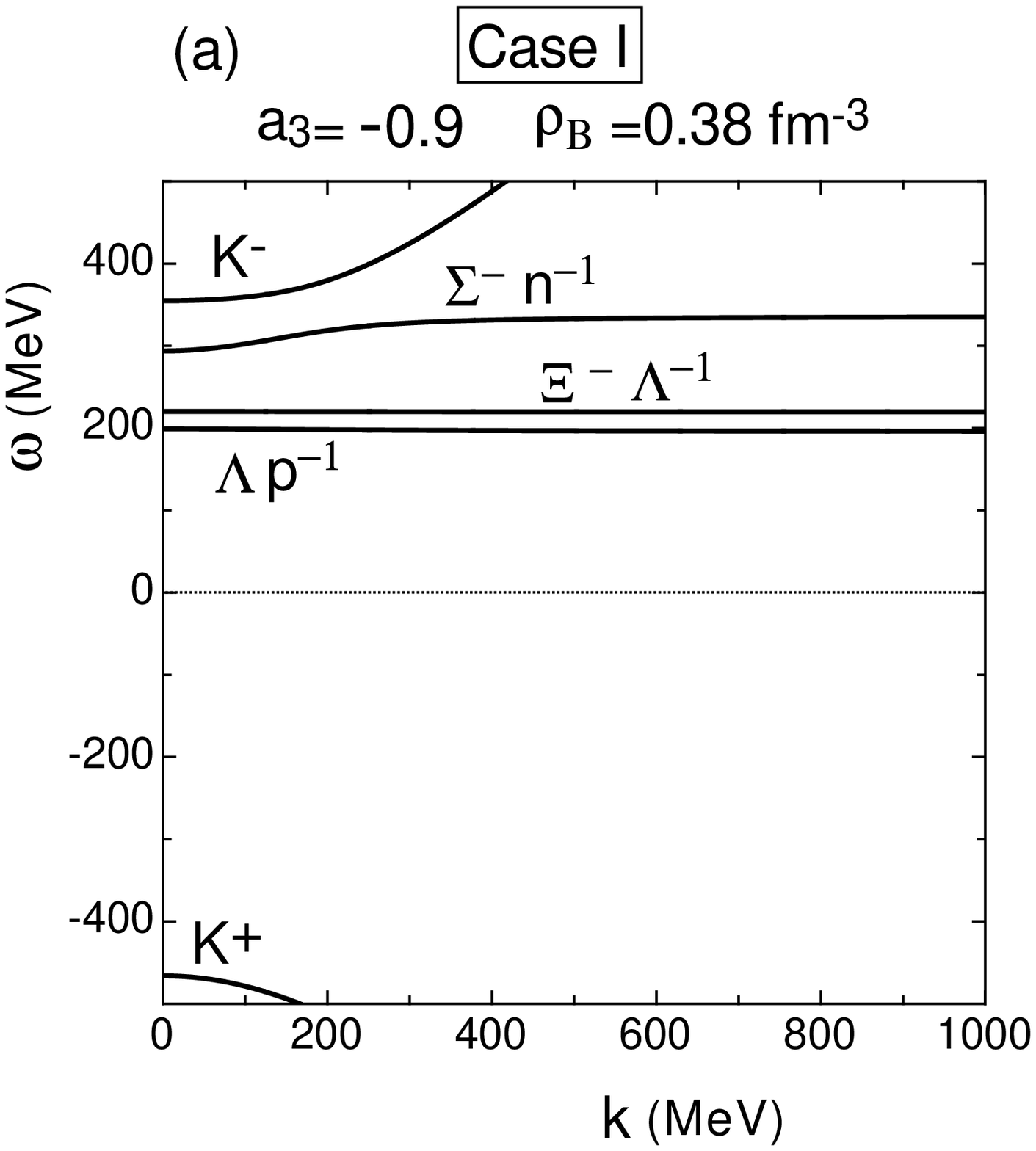}}
\end{minipage}~
\begin{minipage}[r]{0.49\textwidth}
\centerline{
\epsfxsize=\textwidth
\epsffile{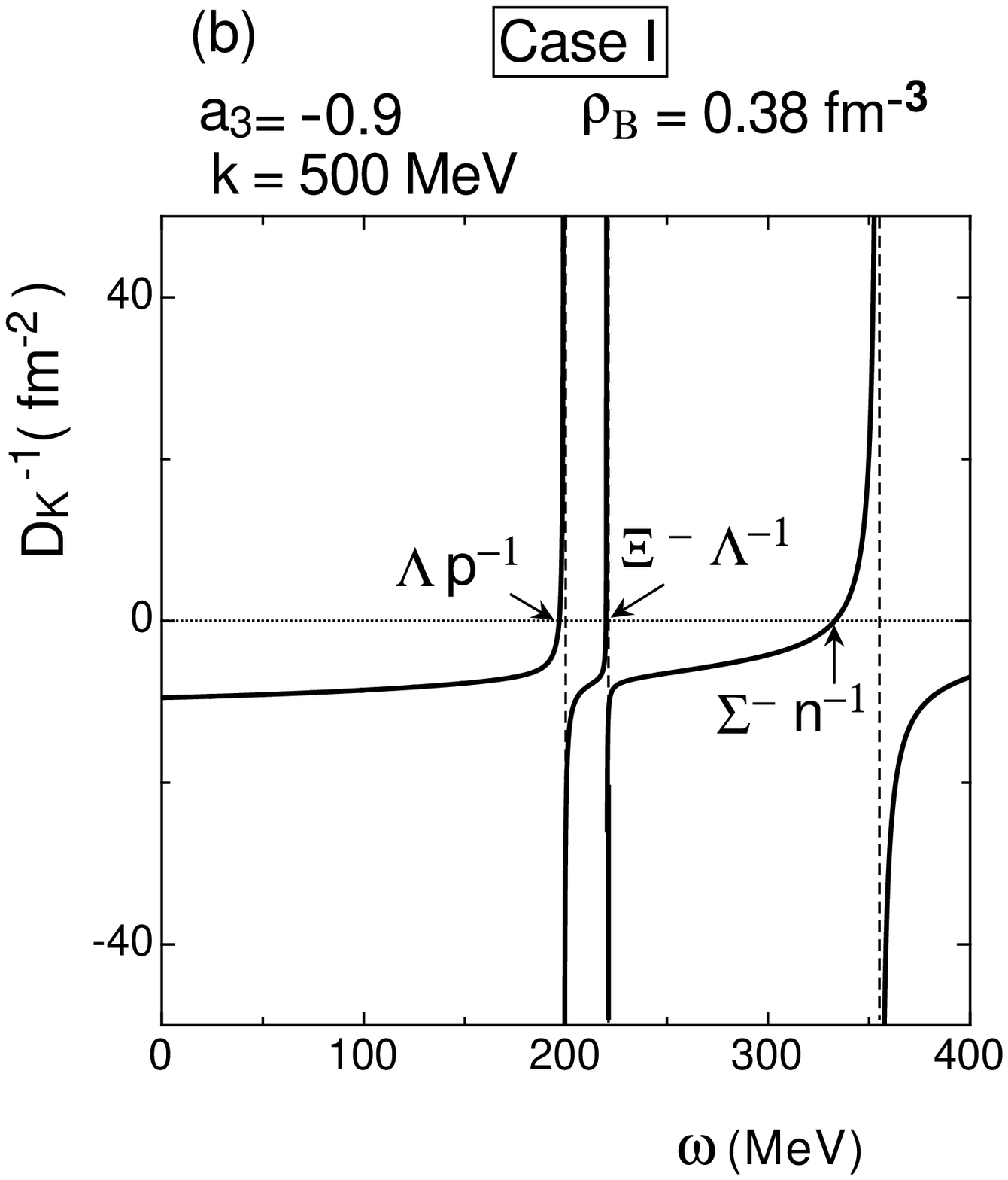}}
\end{minipage}
\caption{(a) The excitation energies for kaonic modes as functions 
of the kaon momentum $|{\bf k}|$ for $a_3=-0.9$ ($\Sigma_{Kn}$=305 
MeV) and the baryon number density $\rho_B$=0.38 fm$^{-3}$.   (b)  
The value of the inverse kaon propagator $D_K^{-1}$ as a function of 
the excitation energy $\omega$ at $|{\bf k}|$=500 MeV for the same 
$a_3$ and density as Fig.~4~(a).}
\label{fig:w-300-038}
\end{figure}

Here we discuss kaonic excitations in hyperonic matter 
in Case I. 
In Fig.~\ref{fig:w-300-038}~(a), we show the excitation energies for  
kaonic modes as functions of the kaon momentum $|{\bf k}|$ for 
$a_3=-0.9$ ($\Sigma_{Kn}$=305 MeV) and $\rho_B$=0.38 fm$^{-3}$ in 
Case I. Around this density,  the $\Lambda$ begins to appear in a 
neutron-star matter [ Fig.\ref{fig:frac}~(a) ], where 
$\rho_\Lambda<\rho_p$. 
In addition to the $K^-$ and $K^+$ branches, 
there are three particle-hole branches: $\Lambda p^{-1}$, $\Xi^- 
\Lambda^{-1}$, and $\Sigma^- n^{-1}$. The excitation modes are 
discriminated by a sign of the residue $\displaystyle (\partial 
D_K^{-1}/\partial \omega)^{-1}$ at their pole of the Green's function 
$D_K$, as is the case with pionic modes\cite{m78,bc79,ew88}: If $ 
\partial D_K^{-1}(\omega,{\bf k};\rho_{\rm B})/\partial \omega >0$, 
the mode has a $K^-$ quantum number, while if $ \partial 
D_K^{-1}(\omega,{\bf k};\rho_{\rm B})/\partial \omega <0$, the mode 
has a $K^+$ quantum number. 
In Fig.~\ref{fig:w-300-038}~(b), the value of the inverse kaon 
propagator $D_K^{-1}$ is shown as a function of the excitation energy 
$\omega$ at $|{\bf k}|$=500 MeV for the same $a_3$ and density as 
Fig.~\ref{fig:w-300-038}~(a). The intersection with the $\omega$ axis 
denotes an excitation energy for each mode. One finds that the three 
particle-hole modes have the $K^-$ quantum numbers at this density. 

As is seen from Fig.~\ref{fig:w-300-038}~(a), the energies of the 
particle-hole branches depends little on the momentum $|{\bf k}|$, 
since the energy for each particle-hole mode is essentially 
determined from the location of the pole in the limit $|{\bf 
k}|\rightarrow 0$ in the $p$-wave part of the self energy 
(Eq.~\ref{eq:selfp}), so long as 
the $p$-wave kaon-baryon coupling strength is not very strong. 
On the other hand, the energies of the $K^+$ and $K^-$ branches are 
sensitive to $|{\bf k}|$. 

\begin{figure}[tt]
\noindent\begin{minipage}[l]{0.50\textwidth}
\centerline{
\epsfxsize=\textwidth
\epsffile{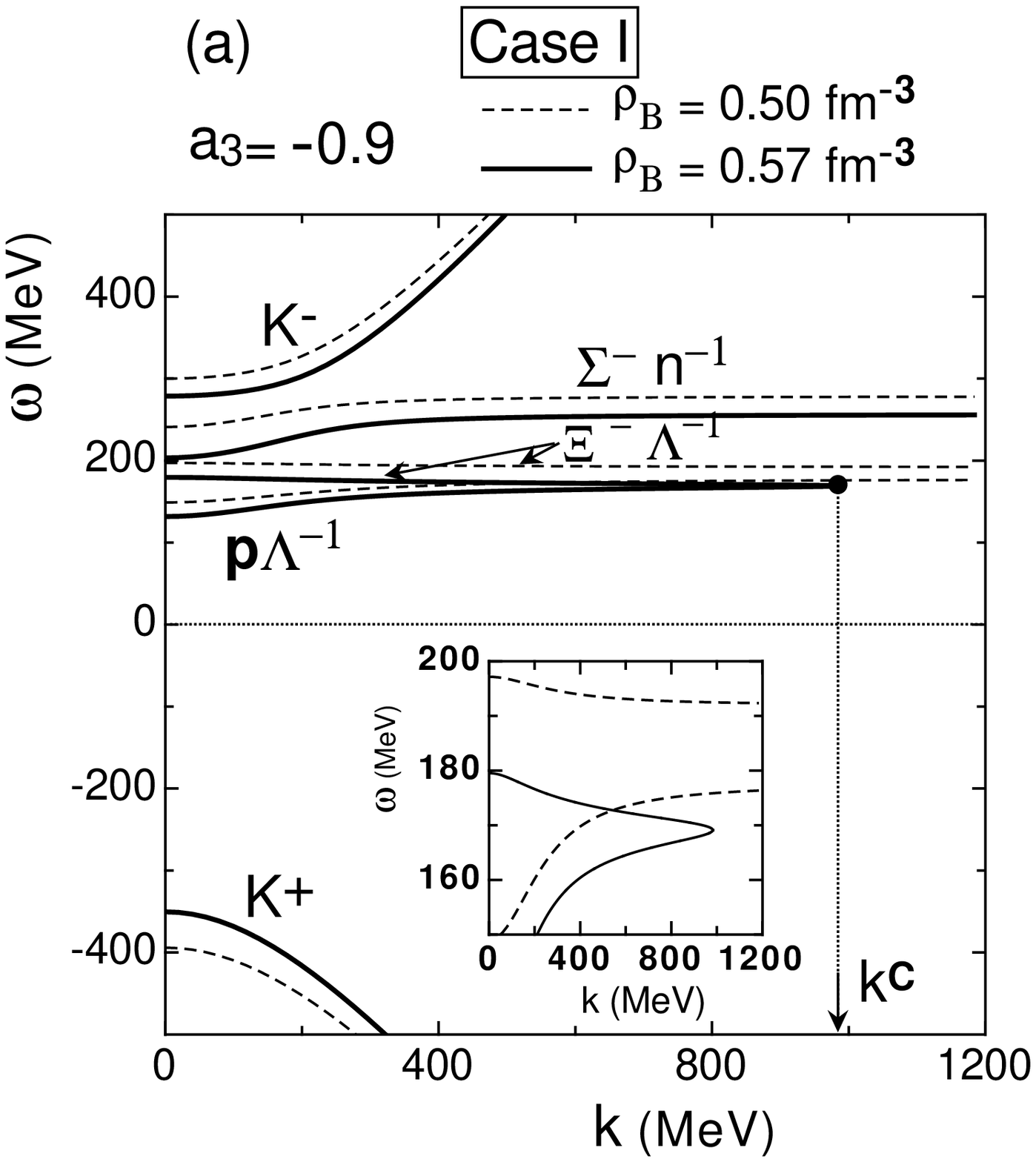}}
\end{minipage}~
\begin{minipage}[r]{0.48\textwidth}
\centerline{
\epsfxsize=\textwidth
\epsffile{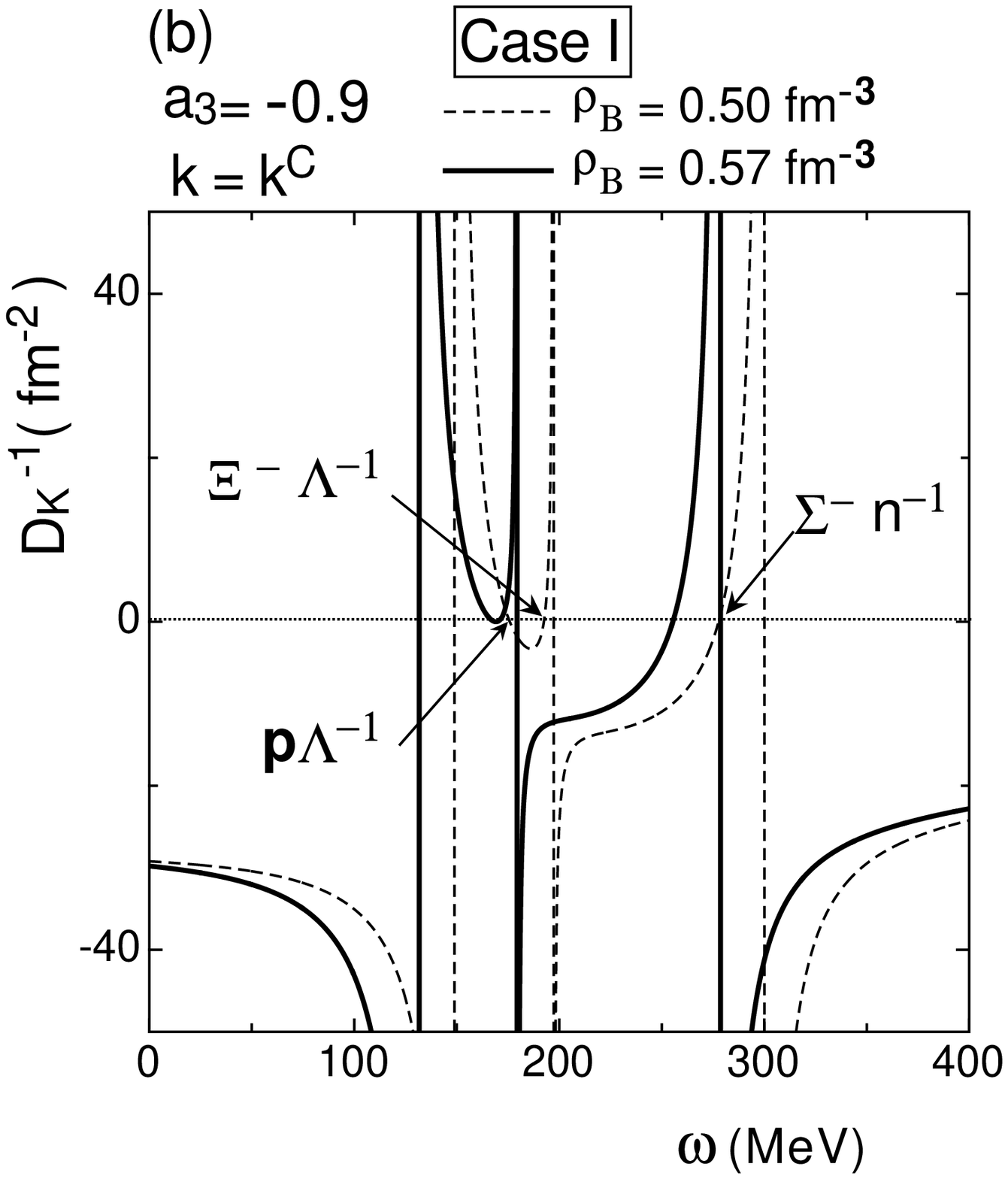}}
\end{minipage}
\caption{(a) The excitation energies for kaonic modes as functions 
of $|{\bf k}|$ for $a_3=-0.9$ and $\rho_{\rm B}$=0.50 fm$^{-3}$ 
(dashed line) and $\rho_{\rm B}$=0.57 fm$^{-3}$ (solid line). 
(b) The inverse kaon propagator $D_K^{-1}(\omega, {\bf k}; \rho_{\rm 
B})$ as a function of $\omega$ at $|{\bf k}|$=$k^{\rm C}$ (=984 MeV) 
for 
the same $a_3$ and  $\rho_{\rm B}$ as in Fig.~5~(a). }
\label{fig:300-rvsm}
\end{figure}

Now we look into the behavior of the excitation modes at the higher 
density where the $\Lambda$ is fully mixed and satisfies 
$\rho_\Lambda>\rho_p$. 
In Fig.~\ref{fig:300-rvsm}~(a), we show the excitation energies for 
kaonic modes as functions of $|{\bf k}|$ for $a_3=-0.9$ and 
$\rho_{\rm B}$=0.50 fm$^{-3}$ (dashed lines) and $\rho_{\rm B}$=0.57 
fm$^{-3}$ (solid lines). The inset in Fig.~\ref{fig:300-rvsm}~(a) 
shows the magnified part of the $p\Lambda^{-1}$ and $\Xi^- 
\Lambda^{-1}$ branches.

In Fig.~\ref{fig:300-rvsm}~(b), 
the value of $D_K^{-1}(\omega, {\bf k}; \rho_{\rm B})$ as a function 
of $\omega$ at certain momentum $|{\bf k}|$=$k^{\rm C}$ (=984 MeV) is 
shown for the same $a_3$ and  $\rho_{\rm B}$ as in 
Fig.~\ref{fig:300-rvsm}~(a). 
For these high densities, the $p\Lambda^{-1}$ branch 
which has a quantum number of the $K^+$ appears instead of the 
$\Lambda p^{-1}$ branch: E.g., for $\rho_{\rm B}$=0.50 fm$^{-3}$, 
$\partial D_K^{-1}/\partial\omega <0$ at the pole of the 
$p\Lambda^{-1}$ mode 
[ Fig.~\ref{fig:300-rvsm}~(b) ]. 
In order to go into details about the condition for the appearance of 
the $p\Lambda^{-1}$ mode, one obtains from Eqs.(\ref{eq:dkinv}) and 
(\ref{eq:self}), 
\begin{eqnarray}
\partial D_K^{-1}/\partial\omega&=& 
2\omega+\frac{1}{f^2}\Bigg\lbrack\Big(
\rho_p+\frac{1}{2}\rho_n-\frac{1}{2}\rho_{\Sigma^-}
-\rho_{\Xi^-} \Big) 
+\frac{1}{2}(\rho_p-\rho_\Lambda)\Bigg(\frac{g_{\Lambda p}{\bf 
k}}{\delta M_{\Lambda p}-\omega+V_\Lambda-V_p}\Bigg)^2 \cr
&+&\frac{1}{2}(\rho_\Lambda-\rho_{\Xi^-})\Bigg(\frac{g_{\Xi^-\Lambda 
}{\bf k}}{\delta M_{\Xi^-\Lambda}-\omega+V_{\Xi^-}-V_\Lambda}\Bigg)^2 
\cr
&+&\frac{1}{2}(\rho_n-\rho_{\Sigma^-})
\Bigg(\frac{g_{\Sigma^- n }{\bf k}}{\delta M_{\Sigma^- 
n}-\omega+V_{\Sigma^-}-V_n}\Bigg)^2 \Bigg\rbrack  \ .  
\label{eq:condition}
\end{eqnarray}
Since the excitation energy $\omega(p\Lambda^{-1})$ for the 
$p\Lambda^{-1}$ mode roughly satisfies $\delta M_{\Lambda 
p}-\omega(p\Lambda^{-1})+V_\Lambda-V_p\sim 0$, the second term in the 
bracket on the r.h.s. of Eq.~(\ref{eq:condition}) coming from the 
$K^-p\Lambda$ interaction is dominant, and a sum of the remaining 
terms in (\ref{eq:condition}) is positive. 
Hence, in order to satisfy $\partial D_K^{-1}/\partial\omega <0$, the 
$\Lambda$ have to be more abundant than the proton ($\rho_\Lambda 
>\rho_p$), which is the necessary 
(but not sufficient) condition for 
the existence of the $p\Lambda^{-1}$ mode. 
In Case I, the $p\Lambda^{-1}$ mode appears at $\rho_{\rm B}\sim$ 
0.40 fm$^{-3}$, where $\rho_p\simeq\rho_\Lambda$ [see 
Fig.~\ref{fig:frac}].  

As the baryon number density increases, the locations of the 
particle-hole branches become lower owing to the $p$-wave 
interactions, as shown in Fig.~\ref{fig:300-rvsm}~(a).  In 
particular, the $\Xi^-\Lambda^{-1}$ and $p\Lambda^{-1}$ branches get 
close to each other, and they merge at certain density 
($\rho_{\rm B}\simeq$ 0.57 fm$^{-3}$) with a 
critical momentum $k^{\rm C}$ (=984 MeV). 
At $|{\bf k}|=k^{\rm C}$,  these two excitation modes merge at the 
$\omega$ axis in the $D_K^{-1}-\omega$ plane 
[Fig.~\ref{fig:300-rvsm}~(b) ], where the double-pole condition, 
\begin{mathletters}\label{eq:double}
\begin{eqnarray}
& & D_{\rm K}^{-1}(\omega, {\bf k} ; \rho_{\rm 
B}) =0 \ , \label{eq:double1} \\
& &\partial D_{\rm K}^{-1}(\omega, {\bf k} ; 
\rho_{\rm B}) /\partial \omega=0  \ , 
\label{eq:double2} 
\end{eqnarray}
\end{mathletters}
and the extremum condition with respect to $|{\bf k}|$, 
\begin{equation}
\partial D_{\rm K}^{-1}(\omega, {\bf k} ; 
\rho_{\rm B}) /\partial |{\bf k}|=0 \ , 
\label{eq:dwdk}
\end{equation}
are satisfied. It means that a pair of the two modes, 
$\Xi^- \Lambda^{-1}$ and $p\Lambda^{-1}$, are created 
spontaneously with no cost of energy because the energy of 
$p\Lambda^{-1}$ mode with the quantum  number $K^+$ is to be reversed 
in sign. Hence the system is unstable with respect to a pair creation 
of [$\Xi^- \Lambda^{-1}$] and [$p\Lambda^{-1}$] modes. This 
instability originates from the $p$-wave kaon-baryon interaction, 
and we call this instability $p$-wave kaon condensation. The onset 
mechanism of the $p$-wave kaon condensation is similar to that of 
pion condensation, where a driving force is given by the $p$-wave 
$\pi N$ interaction\cite{m78,bc79,ew88,kmttt93}. 
From Eqns.~(\ref{eq:double1}), (\ref{eq:double2}), (\ref{eq:dwdk}), 
and by putting $\omega=\mu$ (the charge chemical potential),  one 
obtains the baryon number density $\rho_{\rm B}^{\rm C}$, the charge 
chemical potential $\mu^{\rm C}$, and the kaon momentum $k^{\rm C}$ 
at the onset of condensation. It is to be noted that the numerical value 
of $k^{\rm C}$ obtained for $a_3=-0.9$ in Case I may be very large. 
As a realistic effect, form factors at the $p$-wave kaon-baryon vertices 
might reduce the large $p$-wave attraction,  resulting in a moderate value of 
$k^{\rm C}$, while pushing a critical density $\rho_{\rm B}^{\rm C}$ 
to a higher baryon number density. 

\begin{figure}[t]
\centerline{
\epsfxsize=0.5\textwidth
\epsffile{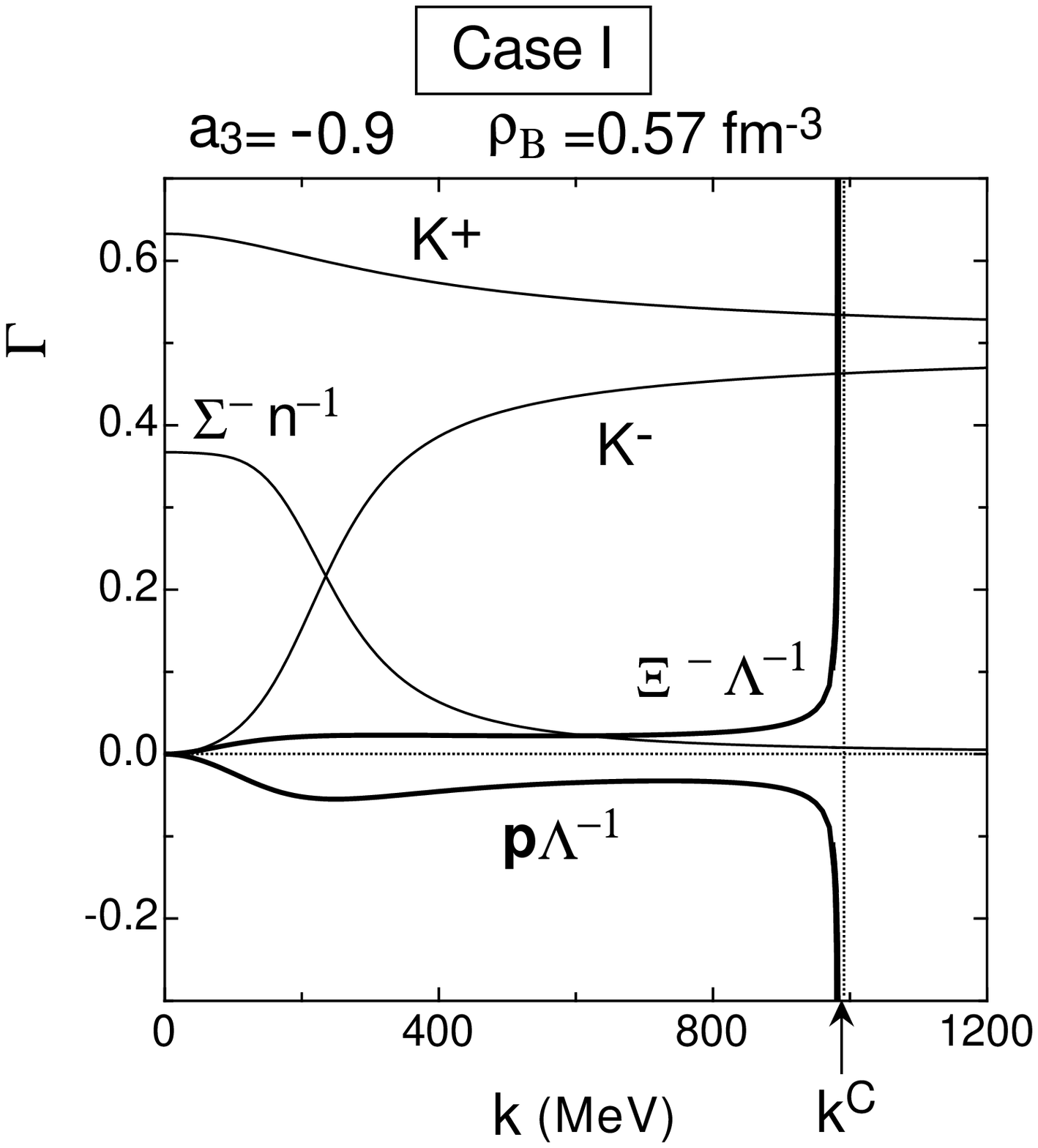}}
\caption{Occupation factors 
$\displaystyle \Gamma (i)\equiv\omega(\partial 
D_K^{-1}/\partial\omega)^{-1}|_{\omega=\omega_i}$ for the kaonic 
modes ($i=K^-, p\Lambda^{-1}, \Sigma^-n^{-1}, 
\Xi^- \Lambda^{-1}, K^+$) as functions of $|{\bf k}|$ for $a_3=-0.9$ 
and $\rho_{\rm B}$=0.57 fm$^{-3}$ in Case I. }
\label{fig:gam1}
\end{figure}

The population of the modes can be seen from the spectral density which is 
defined as $A(\omega, {\bf k}; \rho_{\rm B})=-2  {\rm Im} D^{\rm R}(\omega, {\bf k};
\rho_{\rm B})$, where $D^{\rm R}(\omega, {\bf k}; \rho_{\rm B})$ is the retarded 
Green's function for kaons. In our formulation, the particle-hole continuum states 
are not included, so that the spectral density has a form 
\begin{equation}
A(\omega, {\bf k}; \rho_{\rm B})=2\pi \sum_i Z_i \delta(\omega-\omega_i) \ , 
\label{eq:spectral}
\end{equation}
where $\omega_i$ ($i=K^-, p\Lambda^{-1}, \Sigma^- n^{-1}, \Xi^-\Lambda^{-1}, K^+$) 
are the solutions of the dispersion equation, $D_K^{-1}(\omega, {\bf k}; \rho_{\rm B})=0$, 
and $Z_i$ [$=1/(\partial D_K^{-1}/\partial \omega)_{\omega=\omega_i}$] are the residues 
of $D_K(\omega, {\bf k}; \rho_{\rm B})$ at $\omega=\omega_i$. The spectral density 
has a sum rule, \begin{equation}
 \int_{-\infty}^\infty\frac{d\omega}{2\pi}\omega A(\omega, {\bf k}; 
 \rho_{\rm B})=\sum_i\omega_i Z_i=1 \ , 
 \label{eq:sum}
 \end{equation}
 which follows from the canonical commutation relation for the charged kaon field. 
In Fig.~\ref{fig:gam1}, we show occupation factors\cite{kvk95} which are defined as 
$\displaystyle \Gamma (i)\equiv\omega_iZ_i$  for the 
kaonic modes ($i=K^-, p\Lambda^{-1}, \Sigma^-n^{-1}, \Xi^- 
\Lambda^{-1}, K^+ $) as functions of $|{\bf k}|$ for $a_3=-0.9$ and 
$\rho_{\rm B}$=0.57 fm$^{-3}$ in Case I. The $p\Lambda^{-1}$ mode has a negative  
contribution to the occupation factor. For $|{\bf k}|<k^{\rm C}$, 
the occupation factors for both the $\Xi^-\Lambda^{-1}$ and $p\Lambda^{-1}$ 
are small in comparison with those for the other modes. Near the critical 
momentum $k^{\rm C}$, however, they become large and diverge at $k^{\rm C}$, 
which implies the instability of the system. 

\begin{figure}[t]
\centerline{
\epsfxsize=0.5\textwidth
\epsffile{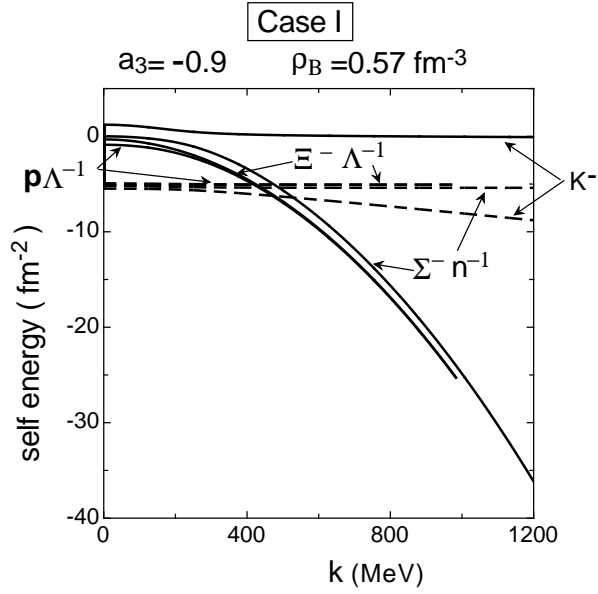}}
\caption{Contributions to the kaon self energy $\Pi_K(\omega, {\bf 
k};\rho)$ from the $p$-wave (solid lines) and $s$-wave (dashed lines)  
kaon-baryon interactions for the kaonic modes as a function of $|{\bf 
k}|$ for $a_3=-0.9 $ and $\rho_{\rm B}$=0.57 fm$^{-3}$ in Case I.  }
\label{fig:self}
\end{figure}

The effects of the $p$-wave kaon-baryon interaction on the 
kaon dynamics near the onset density of condensation are evaluated from 
the self energy $\Pi_K$. 
For each kaonic mode, we show, in Fig.~\ref{fig:self}, the kaon self 
energy for the $s$-wave part $\Pi_K^s(\omega, {\bf k};\rho_{\rm B})$ 
[Eq.~(\ref{eq:selfs})] and the $p$-wave part $\Pi_K^p (\omega, {\bf 
k};\rho_{\rm B})$ [Eq.~(\ref{eq:selfp})] by the dashed lines and the 
solid lines, respectively, as a function of $|{\bf k}|$ for $a_3=-$ 
0.9 and $\rho_{\rm B}=\rho_{\rm B}^{\rm C}$=0.57 fm$^{-3}$. 
For the $p\Lambda^{-1}$, $\Sigma^-n^{-1}$, and $\Xi^- \Lambda^{-1}$, 
the attractive $p$-wave part $\Pi_K^p$ gets large almost 
proportionally to $|{\bf k}|^2$, and the magnitude becomes comparable 
to that of the $s$-wave part $\Pi_K^s$ at $|{\bf k}|\sim$ 500 MeV. 

 On the other hand, the $p$-wave part $\Pi_K^p$ for the $K^-$ 
works  repulsively at small $|{\bf k}|$, and it decreases 
monotonically 
with $|{\bf k}|$. At a high momentum, the excitation energy for the 
$K^-$ mode is so large that the  $K^-$ mode is located far beyond the 
poles for the other particle-hole modes, which yields 
$\delta M_{\Lambda p}-\omega(K^-)+V_\Lambda-V_p\ll 0$, 
$\delta M_{\Sigma^- n}-\omega(K^-)+V_{\Sigma^-}-V_n\ll 0$, 
and $\delta M_{\Xi^-\Lambda}-\omega(K^-)+V_{\Xi^-}-V_\Lambda \ll 0$ 
in the $\Pi_K^p$. Hence the magnitude of the $p$-wave part $\Pi_K^p$ 
for the $K^-$ is tiny, 
and the $s$-wave part $\Pi_K^s$ is dominant in the self energy. 
\vspace{0.5cm}

\noindent\underline{(ii) The stronger $s$-wave scalar interaction 
($a_3=-1.1$)}
\vspace{0.3cm}

\begin{figure}[t]
\noindent\begin{minipage}[l]{0.50\textwidth}
\centerline{
\epsfxsize=\textwidth\epsffile{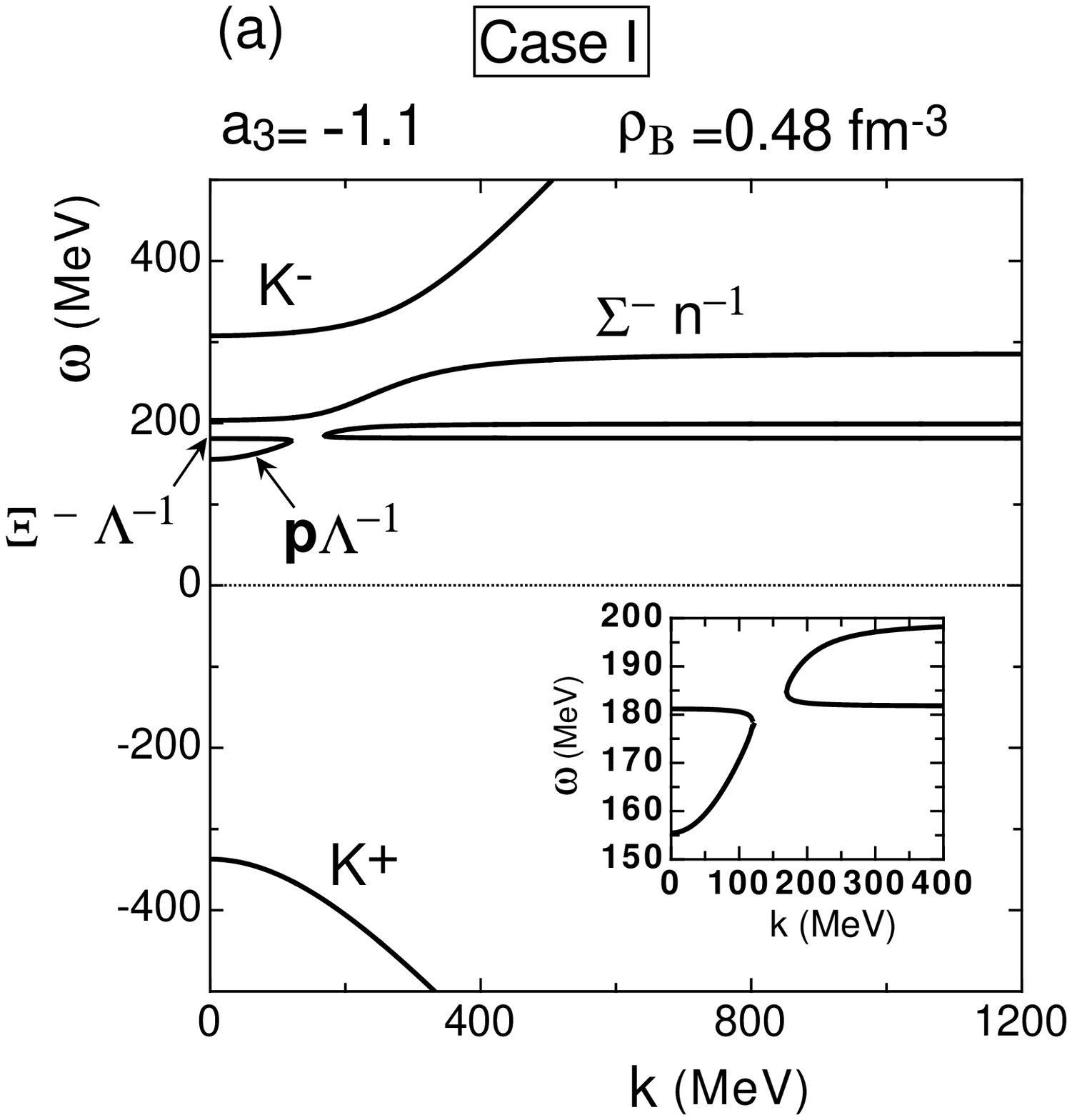}}
\end{minipage}~
\begin{minipage}[r]{0.46\textwidth}
\centerline{
\epsfxsize=\textwidth\epsffile{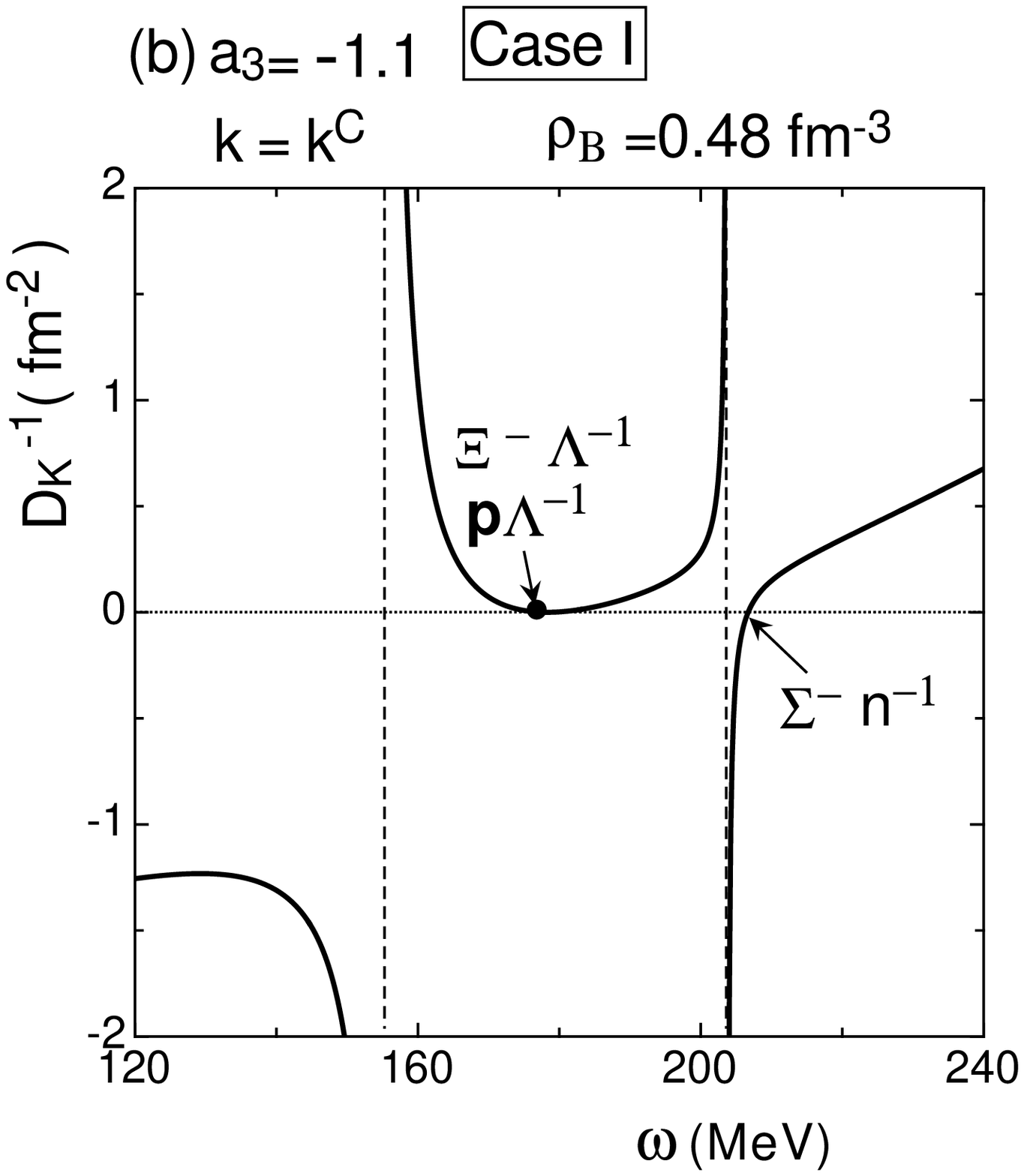}}
\end{minipage}
\caption{(a) The excitation energies of kaonic modes as functions  of 
$|{\bf k}|$ for $a_3=-1.1$ ($\Sigma_{Kn}$=403 MeV) 
and $\rho_{\rm B}$=0.48 fm$^{-3}$ just beyond the onset of 
condensation in Case I. (b) The value of the kaon inverse propagator 
$D_K^{-1}(\omega,{\bf k}; \rho_{\rm B})$ as a function of $\omega$ 
at $|{\bf k}|=k^{\rm C}$=118 MeV. The values of $a_3$ and $\rho_B$ 
are the same as those in Fig.~8~(a).  }
\label{fig:400-rvsm}
\end{figure}

\begin{figure}[h]
\noindent\begin{minipage}[l]{0.50\textwidth}
\centerline{
\epsfxsize=\textwidth\epsffile{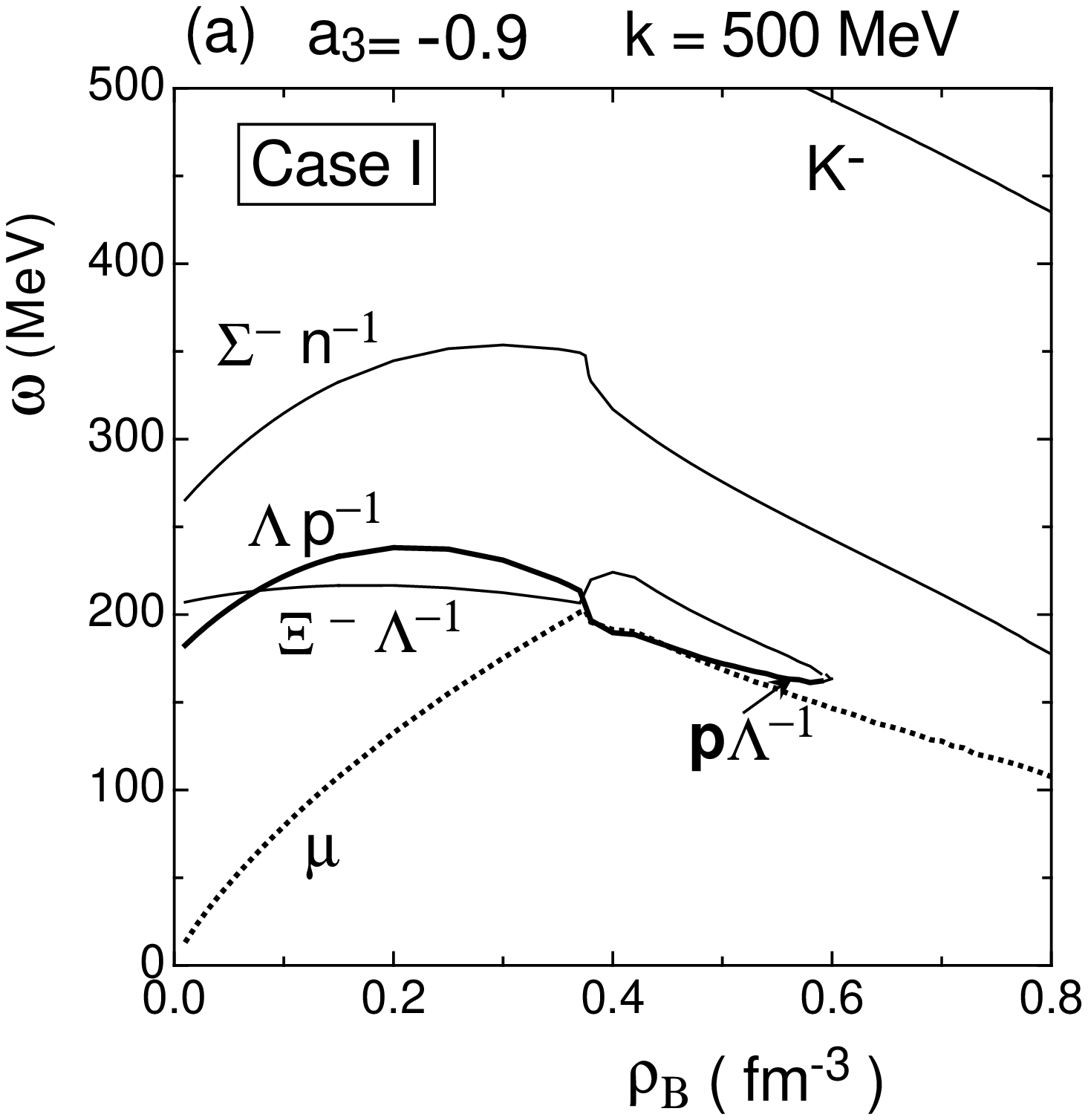}}
\end{minipage}~
\begin{minipage}[r]{0.50\textwidth}
\centerline{
\epsfxsize=\textwidth\epsffile{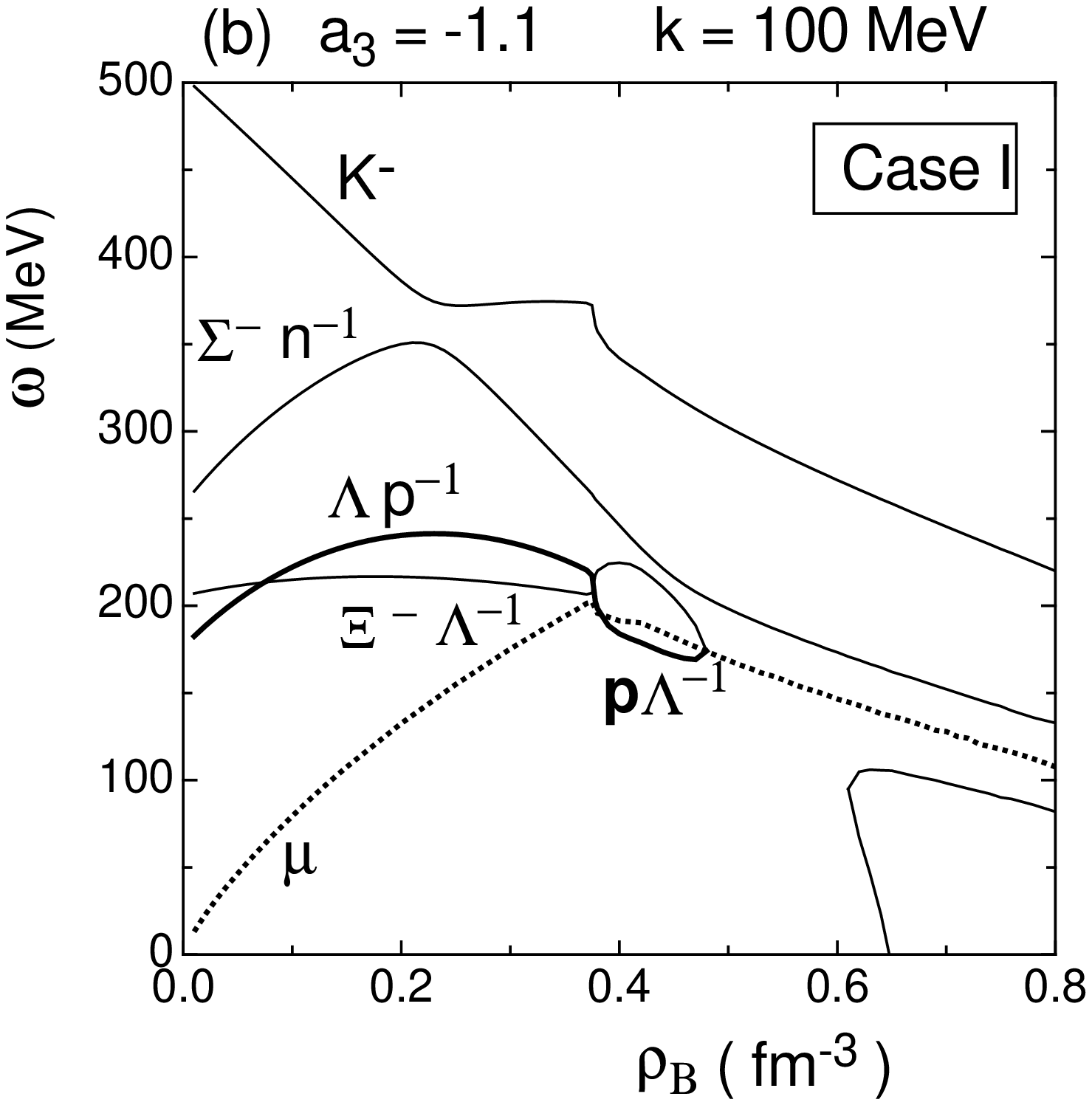}}
\end{minipage}
\caption{(a) The dependence of the excitation energies of kaonic 
modes on the baryon number density for $a_3=-0.9$ ($\Sigma_{Kn}$=305 
MeV) and 
$|{\bf k}|$=500 MeV in Case I. The charge chemical potential $\mu$ is 
also shown as a function of $\rho_{\rm B}$ by a dotted line. 
(b) The same as Fig.~9~(a), but for  $a_3=-1.1$ ($\Sigma_{Kn}$=403 
MeV) and $|{\bf k}|$=100 MeV. The charge chemical potential $\mu$ 
(the dotted line) is identical to that in Fig.~9~(a). }
\label{fig:w-rho-rvsm}
\end{figure}

Next we consider a case for the stronger $s$-wave scalar attraction. 
In Fig.~\ref{fig:400-rvsm}~(a), we show 
the excitation energies of kaonic modes as functions of 
$|{\bf k}|$ for $a_3=-1.1$ ($\Sigma_{Kn}$=403 MeV) 
and $\rho_{\rm B}$=0.48 fm$^{-3}$ just beyond the onset of 
condensation. In Fig.~\ref{fig:400-rvsm}~(b), the value of the kaon 
inverse propagator 
$D_K^{-1}(\omega,{\bf k}; \rho_{\rm B})$ is shown as a function of 
$\omega$ at $|{\bf k}|=k^{\rm C}$=118 MeV for the same values of 
$a_3$ and $\rho_B$ as those in Fig.~\ref{fig:400-rvsm}~(a). 
As is the case with $a_3=-0.9$, the $p$-wave condensation of the 
[$\Xi^-\Lambda^{-1}$] and [$p\Lambda^{-1}$] pairs occurs but at a 
smaller density $\rho_{\rm B}$=0.48 fm$^{-3}$ and a smaller momentum 
$|{\bf k}|^{\rm C}$=118 MeV than those for $a_3=-0.9$.  The  
difference of the critical density and the critical momentum between 
the two cases is attributed to the difference of the microscopic 
structures of the kaonic modes. 
 Figure~\ref{fig:400-rvsm}~(a) shows that there are level crossings 
between the $K^-$, $\Sigma^- n^{-1}$ and $\Xi^- \Lambda^{-1}$ 
branches, whereas there is a level crossing only between the $K^-$ 
and $\Sigma^-n^{-1}$ branches for $a_3=-0.9$ 
[Fig.~\ref{fig:300-rvsm}~(a)].  The difference of the structures for 
the kaonic modes between the $a_3=-0.9$ and $-1.1$ cases can also be 
seen from the dependence of the kaonic modes on the baryon number 
density. In 
Fig.~\ref{fig:w-rho-rvsm}, we show the dependence of the excitation 
energies of kaonic modes on the baryon number density except for the 
$K^+$. (a) is for $a_3=-0.9$ and $|{\bf k}|$=500 MeV which is smaller 
than $k^{\rm C}$ (=984 MeV), and (b) is for $a_3=-1.1$ and $|{\bf 
k}|$=100 MeV, which is smaller than but near $k^{\rm C}$ (=118 MeV). 
For $a_3=-0.9$, the $K^-$ is repelled far from the remaining 
particle-hole modes over the relevant densities, and there is no 
level crossing. The $\Xi^-\Lambda^{-1}$ and $p\Lambda^{-1}$ modes 
merge at $\rho_{\rm B}\simeq$ 0.60 fm$^{-3}$, which corresponds to 
the instability with respect to $p$-wave condensation. The 
qualitative feature is also applied to the case at  the critical 
momentum $k^{\rm C}$ which satisfies the conditions 
Eqs.~(\ref{eq:double}) and (\ref{eq:dwdk}). The particle-hole 
branches such as the $\Xi^-\Lambda^{-1}$ and $p\Lambda^{-1}$ do not 
depend much on the value of $|{\bf k}|$, nor does the critical point 
for the $p$-wave condensation. 
 
 For $a_3=-1.1$, on the other hand, the excitation energy of the 
$K^-$ is small as compared with the $a_3=-0.9$ at a given density due 
to the larger $s$-wave scalar attraction and the smaller momentum 
$|{\bf k}|$, and one can see in Fig.~\ref{fig:w-rho-rvsm}~(b) that 
there are energy gaps between  $K^-$ and $\Sigma^- n^{-1}$ branches 
and the $\Sigma^- n^{-1}$ and $\Xi^-\Lambda^{-1}$ branches owing to 
the level crossings.\footnote{The appearance of the collective modes 
was also pointed out in relation to the level crossing in  
Ref.~\cite{ynm93}. }
As a result of the level crossings, the $\Xi^-\Lambda^{-1}$ branch 
takes over the characteristics of the $K^-$, the excitation energy of 
which changes appreciably depending on the magnitude of the $s$-wave 
scalar interaction simulated by $a_3$.\footnote{This mode corresponds 
to the kaesobar\cite{llb97,blr98}. } Thus, when the level 
crossing occurs, the behavior of the $\Xi^-\Lambda^{-1}$ branch is 
sensitive to the value of $a_3$, and  the critical point for the 
$p$-wave condensation, which is given by the merge point of the 
$\Xi^-\Lambda^{-1}$ and $p\Lambda^{-1}$ branches, also depends 
on  $a_3$. This mechanism of $p$-wave condensation is similar to 
that of $p$-wave pion condensation, where the $\pi^-$ mode, which 
reduces to a free $\pi^-$ in vacuum, and 
the spin-isospin zero sound (called $\pi_s^+$ ) are spontaneously 
created in pairs\cite{m78,bc79,ew88,kmttt93}.

The larger mixing of the $\Lambda$ than the proton in hyperonic 
matter is a necessary condition for the appearance of the low-lying 
particle-hole mode ($p\Lambda^{-1}$) having a $K^+$ quantum number, 
and this condition is crucial to the realization of $p$-wave kaon condensation 
considered here. The result should be compared with a mechanism of $p$-wave 
kaon condensation realized from the conventional neutron-star matter where 
only the nucleons $n$, $p$ are present as baryons\cite{m93}. In the latter case, 
there is no {\it low-lying} collective mode having the $K^+$ quantum number 
in the ground state of the neutron-star matter, so that condensation 
of pair modes with the $K^+$ and $K^-$ quantum numbers cannot be expected. 
Instead, it is a single mode having the $K^-$ quantum number that is relevant 
to kaon condensation in the (nucleonic)  neutron-star matter\cite{m93,kvk95}: 
In particular, it has been shown in Ref.~\cite{m93} that a Bose-Einstein condensation 
of the $s$-wave $K^-$ mode occurs at some density as a result of the $s$-wave 
$K^- N$ attraction, where hyperons are still absent in matter. 
It has also been shown that only at higher densities, the classical $K^-$ field 
simply acquires a momentum, increasing the attractive energy through the $p$-wave 
kaon- baryon interaction in addition to the $s$-wave one, accompanying hyperon 
excitation\cite{m93}. 

\begin{figure}[t]
\centerline{
\epsfxsize=0.5\textwidth\epsffile{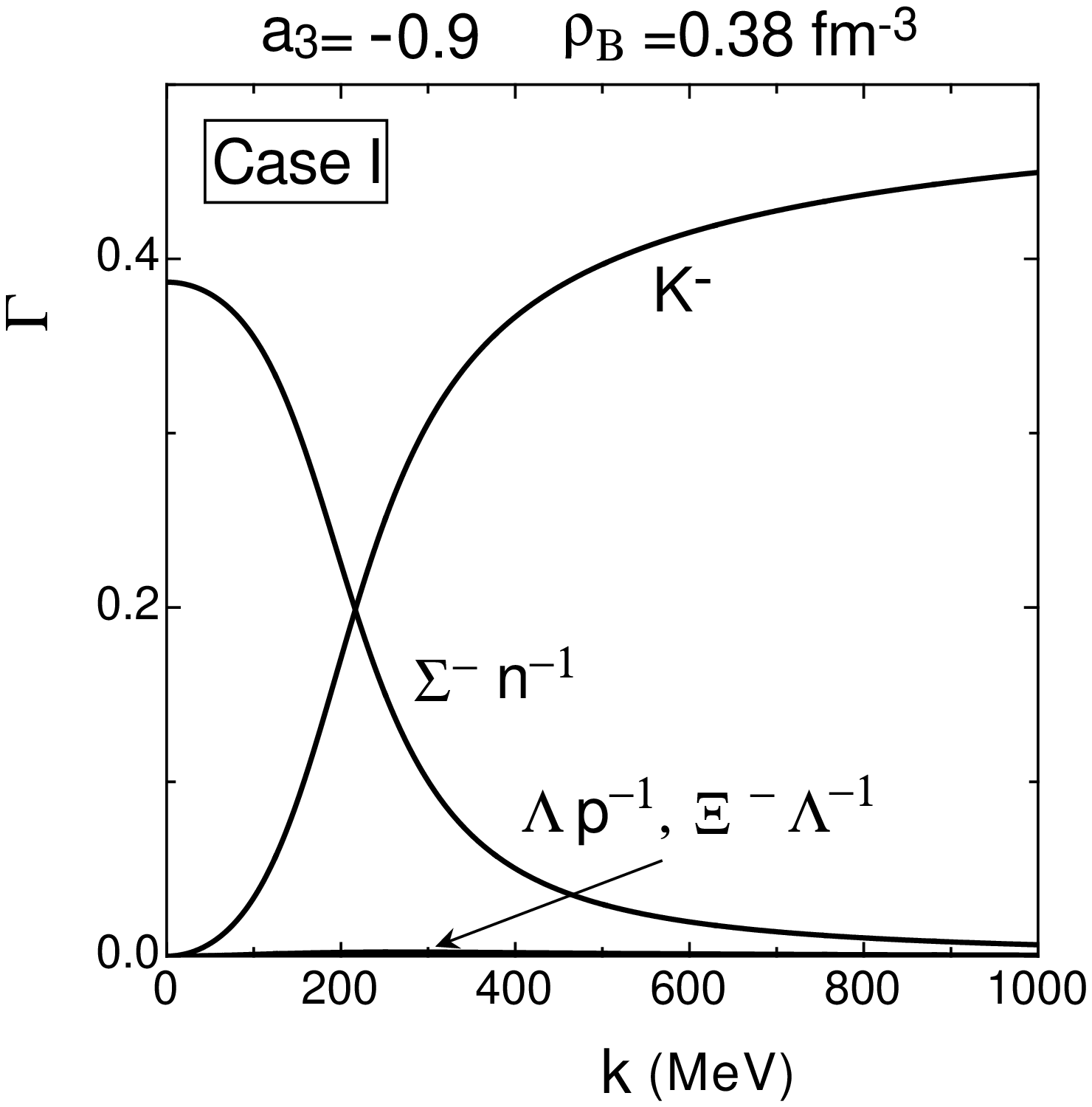}}
\caption{Occupation factors 
$\displaystyle \Gamma (i)\equiv\omega(\partial 
D_K^{-1}/\partial\omega)^{-1}|_{\omega=\omega_i}$ for the kaonic 
modes ($i=K^-, \Lambda p^{-1}, \Sigma^-n^{-1}, 
\Xi^- \Lambda^{-1}$) as functions of $|{\bf k}|$ for $a_3=-0.9$ 
and $\rho_{\rm B}$=0.38 fm$^{-3}$ in Case I. }
\label{fig:gam}
\end{figure}

As seen in Fig.~\ref{fig:w-rho-rvsm}, the $\Lambda p^{-1}$ branch 
crosses the charge chemical potential, 
i. e.,  $\omega(\Lambda p^{-1})=\mu$ at $\rho_{\rm B}\sim$ 0.38 
fm$^{-3}$. It apparently 
suggests an onset of another type of Bose-Einstein  condensation of 
$\Lambda p^{-1}$ mode. In Fig.~\ref{fig:gam}, we show the 
occupation factors $\Gamma(i)$  as functions of $|{\bf k}|$ for $a_3=-0.9$ and 
$\rho_{\rm B}$=0.38 fm$^{-3}$ in Case I. The value of $\Gamma$ for 
the $\Lambda p^{-1}$ mode as well as the $\Xi^-\Lambda^{-1}$ is 
negligible over the relevant kaon momentum $|{\bf k}|$. This is explained as follows: 
The excitation energy for the $\Lambda p^{-1}$ mode is determined 
from $\delta M_{\Lambda p}-\omega(\Lambda p^{-1}) +V_\Lambda-V_p\sim 0$, 
and it depends little on the momentum ${\bf k}$ [ see Fig.~\ref{fig:w-300-038}~(a) 
and the discussion at the beginning of Sec.~\ref{subsubsec:disprvsm}-(i)].  
In this case, the second term in the bracket on the r.h.s. of Eq.~(\ref{eq:condition}) 
for the $\partial D_K^{-1}/\partial\omega$ becomes very large at 
$\omega=\omega(\Lambda p^{-1})$, so that $\Gamma(\Lambda p^{-1})$ 
[ $\propto 1/(\partial D_K^{-1}/\partial\omega)_{\omega=\omega(\Lambda p^{-1})}$] 
becomes very small. The small population of the $\Xi^-\Lambda^{-1}$ is also explained 
in a similar way. The main population of kaonic modes is thus exchanged between 
the $K^-$ and $\Sigma^- n^{-1}$ modes before and after an avoided level-crossing point 
$|{\bf k}|\sim 200 $ MeV [ see Fig.~\ref{fig:w-300-038}~(a) ] .  On the other hand, 
the number of $K^-$ with ${\bf k}$ is given as 
$\displaystyle
 n({\bf k})=\int \frac{d \omega}{2\pi}\omega f_K(\omega)A(\omega, {\bf k}; \rho_{\rm B})=\sum_i f_K(\omega_i)\Gamma(i)  
$
, where $f_K(\omega)$ (=$1/[\exp\{(\omega-\mu)/T\}-1]$) is the Bose-Einstein 
distribution function. Since the $f_K(\omega)$ becomes divergent at 
$\omega=\omega(\Lambda p^{-1})$, the $\Lambda p^{-1}$ mode brings about 
a singular behavior of $n({\bf k})$, however small the population 
$\Gamma(\Lambda p^{-1})$ is. In this respect, the physical significance 
of this instability should be considered carefully. In this paper,
however, we don't go into details about this possible instability, 
and only concentrate on the characteristic features of the pair-mode condensation. 
\subsubsection{Case II (the attractive $V_{\Sigma^-}$)} 
\label{subsubsec:caseII}
\vspace{0.3cm}

\begin{figure}[t]
\noindent\begin{minipage}[l]{0.50\textwidth}
\centerline{
\epsfxsize=\textwidth
\epsffile{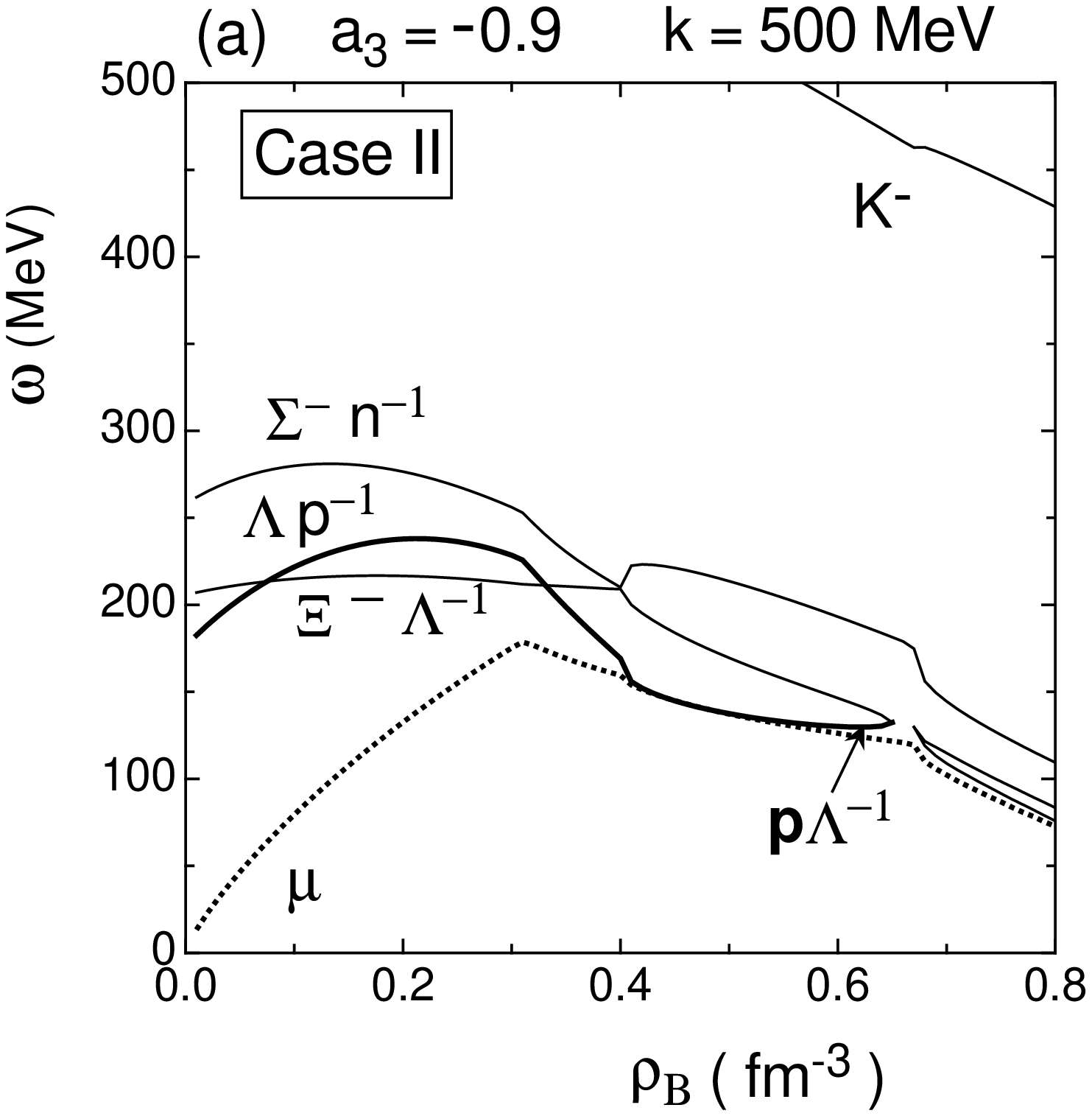}}
\end{minipage}~
\begin{minipage}[r]{0.50\textwidth}
\centerline{
\epsfxsize=\textwidth
\epsffile{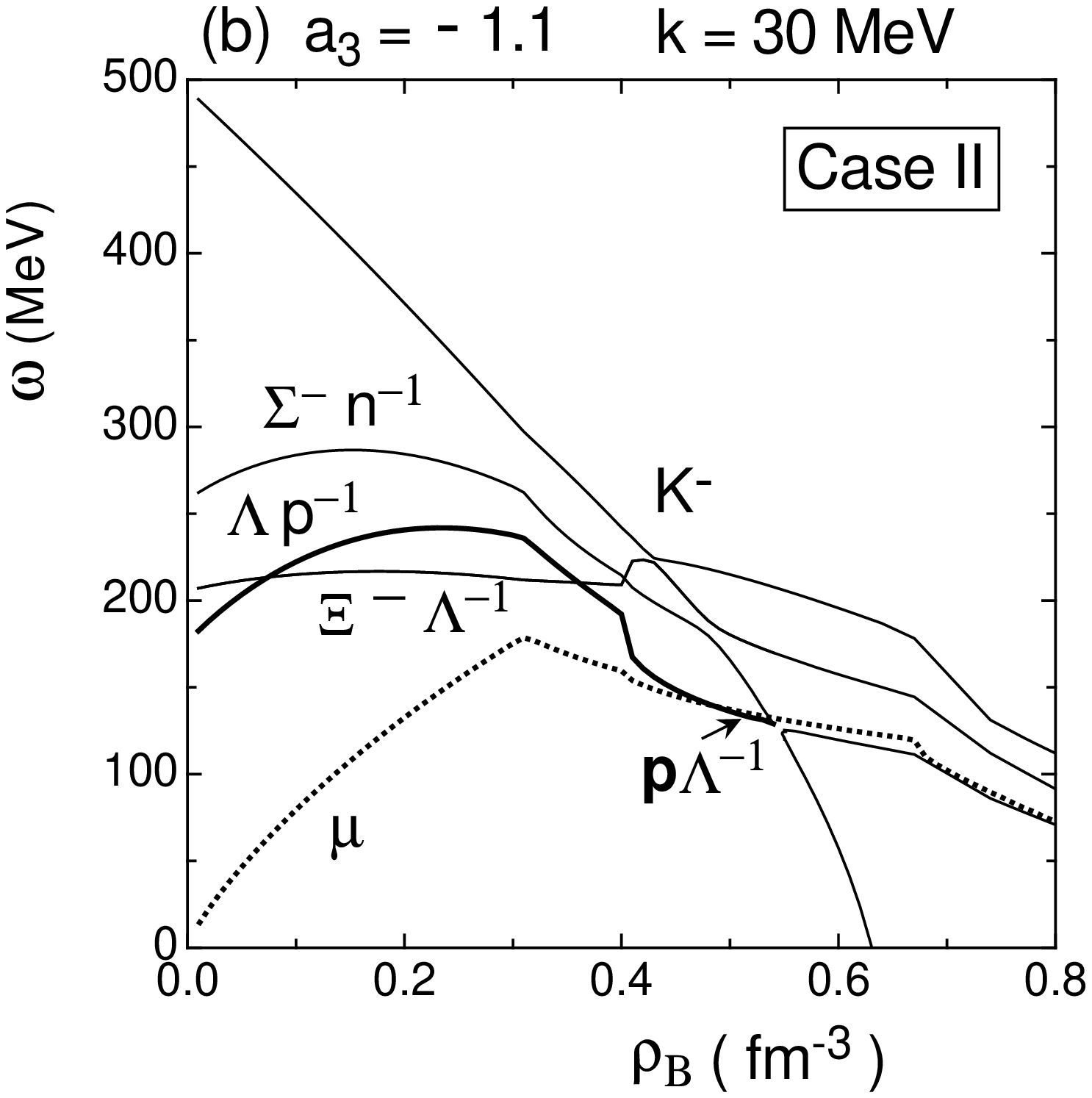}}
\end{minipage}
\caption{(a) The dependence of the excitation energies of kaonic 
modes on the baryon number density for $a_3=-0.9$ ($\Sigma_{Kn}$=305 
MeV) and $|{\bf k}|$=500 MeV in Case II. The charge chemical 
potential $\mu$ is also shown as a function of $\rho_{\rm B}$ by a 
dotted line. (b) The same as Fig.~11~(a), but for  $a_3=-1.1$ 
($\Sigma_{Kn}$=403 
MeV) and $|{\bf k}|$=30 MeV. The charge chemical potential $\mu$ 
(the dotted line) is identical to that in Fig.~11~(a). }
\label{fig:w-rho-avsm}
\end{figure}

Next we discuss behaviors of kaonic modes in Case II (the attractive 
$V_{\Sigma^-}$). In Fig.~\ref{fig:w-rho-avsm}, we show the dependence 
of the excitation energies of kaonic modes on the baryon number 
density in Case II.   (a) is for $a_3=-0.9$ ($\Sigma_{Kn}$=305 MeV) 
and $|{\bf k}|$=500 MeV, and (b) is for $a_3=-1.1$ ($\Sigma_{Kn}$=403 
MeV) and $|{\bf k}|$=30 MeV. 
Due to the strong attraction 
of $V_{\Sigma^-}$,  the $\Sigma^- n^{-1}$ branch is softer than 
those of the $\Xi^-\Lambda^{-1}$ and $K^-$ branches for $\rho_{\rm 
B}\gtrsim$0.40 fm$^{-3}$, and the $\Sigma^- n^{-1}$ merges first with 
the $p\Lambda^{-1}$ branch instead of the $\Xi^-\Lambda^{-1}$. Hence, 
in Case II, $p$-wave condensation is brought about by a 
spontaneous creation of the $\Sigma^- n^{-1}$ and $p\Lambda^{-1}$ 
pair. For the larger $a_3$ [Fig.~\ref{fig:w-rho-avsm}~(b)], there are 
level crossings between the $K^-$, $\Sigma^- n^{-1}$ and 
$\Xi^-\Lambda^{-1}$. The $\Sigma^- n^{-1}$ branch takes over the 
characteristics of the $K^-$ around the crossing  point with the 
$\Xi^-\Lambda^{-1}$ branch [$\rho_{\rm B}\sim$0.53 fm$^{-3}$], so that the 
critical point for the $p$-wave condensation is sensitive to the 
value of $a_3$, which is similar to the stronger $s$-wave attraction 
case ($a_3=-1.1$) in Case I [see Fig.~\ref{fig:w-rho-rvsm}~(b)]. 

The critical density and the corresponding momentum are given from 
Eq.~(\ref{eq:dwdk}) in addition to the double-pole 
condition Eq.~(\ref{eq:double}) as 
$\rho_{\rm B}^{\rm C}$=0.64 fm$^{-3}$, $k^{\rm C}$=978 MeV for 
$a_3=-0.9$, and $\rho_{\rm B}^{\rm C}$=0.53 fm$^{-3}$, $k^{\rm C}$=39 
MeV for $a_3=-1.1$. Quantitatively, the critical density in Case II 
is a little larger than that in Case I. 
As seen from Fig.~\ref{fig:w-rho-rvsm} and Fig.~\ref{fig:w-rho-avsm}, 
the strongly attractive $V_{\Sigma^-}$ in Case II modifies the density 
dependence of the relevant kaonic modes, especially $\Sigma^- n^{-1}$ 
and $p \Lambda^{-1}$ ($\Lambda p^{-1}$), from that in Case I 
( the repulsive $V_{\Sigma^-}$) through changing chemical composition 
of highly dense matter. The quantitative estimation of the density at 
which the two modes merge is subtle depending on the specific density 
dependence of these modes. 
\vspace{0.5cm}

\subsubsection{Comparison of the critical densities for $p$-wave 
and $s$-wave condensations}
\label{subsubsec:ps}
\vspace{0.3cm}

\begin{figure}[t]
\noindent\begin{minipage}[l]{0.5\textwidth}
\centerline{
\epsfxsize=\textwidth
\epsffile{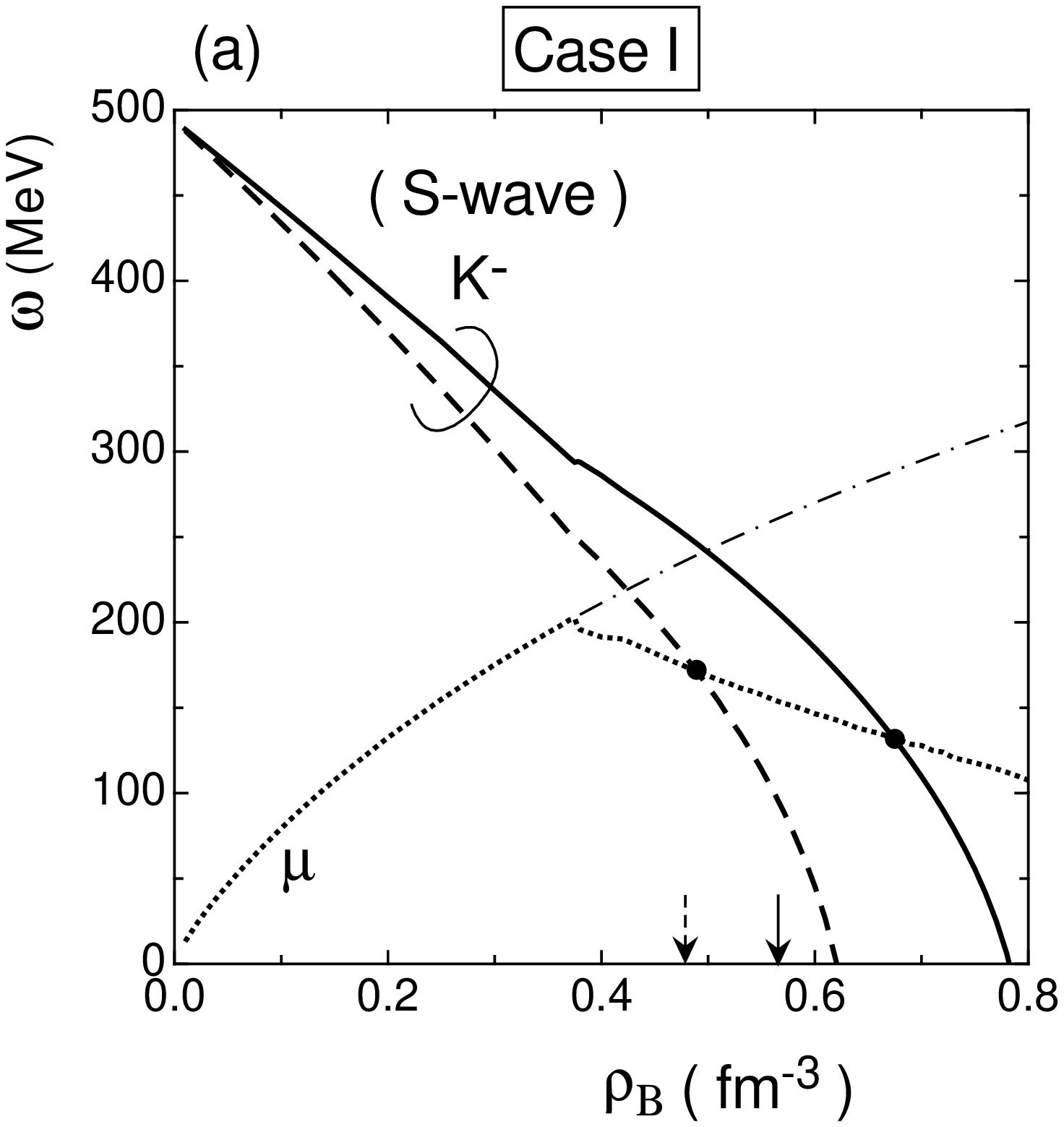}}
\end{minipage}~
\begin{minipage}[r]{0.50\textwidth}
\centerline{
\epsfxsize=\textwidth
\epsffile{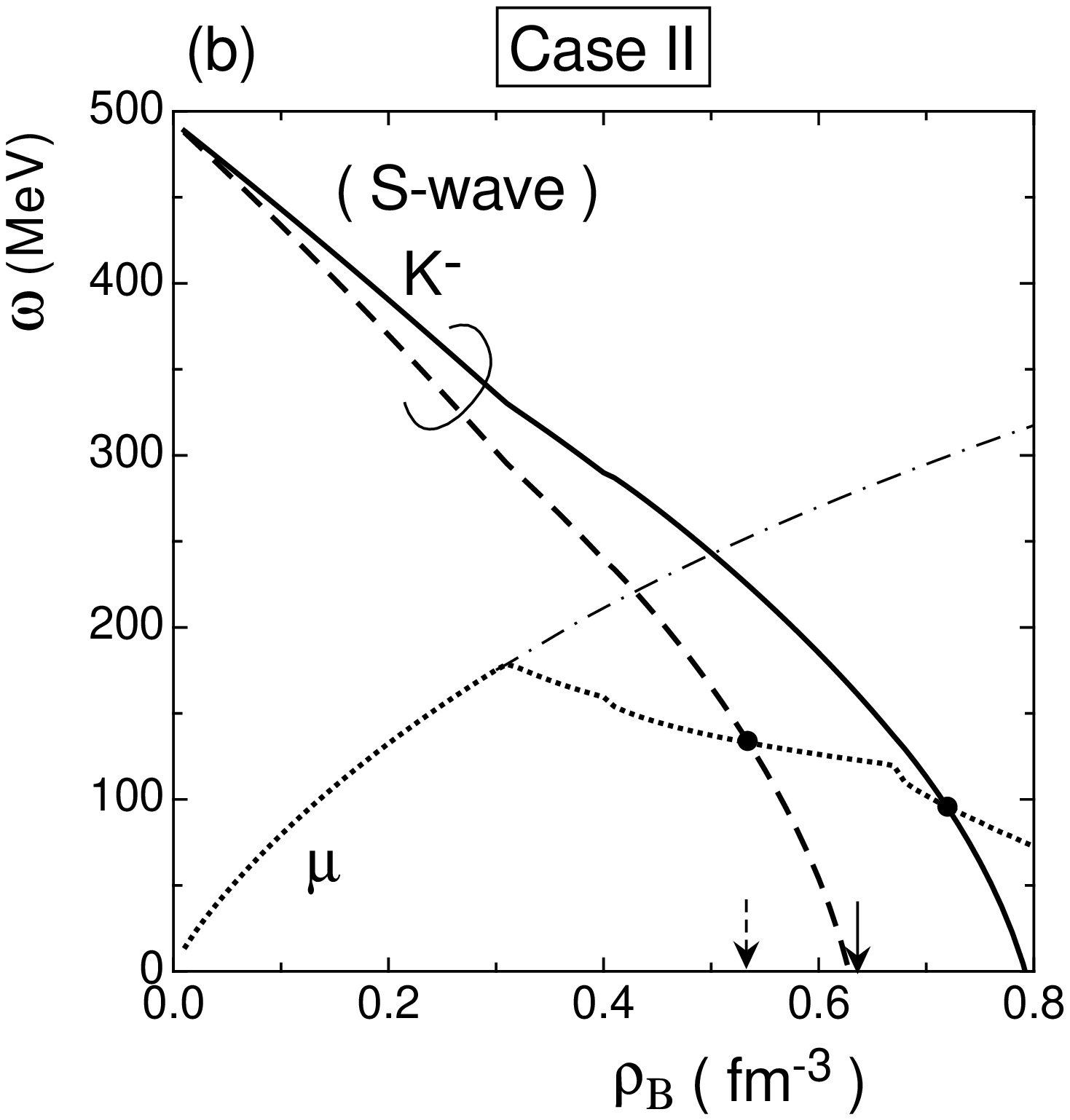}}
\end{minipage}
\caption{(a) The dependence of the minimum 
excitation energy of the $s$-wave $K^-$ on the baryon number density 
for $|{\bf k}|=0$ in Case I. 
The solid line is for $a_3=-0.9$ and the dashed line is for 
$a_3=-1.1$. The charge chemical potential $\mu$ (the dotted line) is 
identical to that in Fig.~9. For reference, the charge chemical 
potential in neutron-star matter consisting of only the nucleons 
$n$, $p$ and $e^-$ is shown by the dash-dotted line.  
(b) The same as Fig.~12 (a), but for Case II. The charge chemical potential 
$\mu$ (the dotted line) is 
identical to that in Fig.~11.  }
\label{fig:swave}
\end{figure}

Here we compare the critical density for the $p$-wave  condensation 
discussed in the preceding subsections with that
for the $s$-wave $K^-$ condensation which is obtained within the 
present framework. In Fig.~\ref{fig:swave}, we show 
the dependence of the minimum excitation energy 
of the $s$-wave $K^-$ on the baryon number density for $|{\bf k}|=0$. 
(a) is for Case I, and (b) is for Case II. The solid line is 
for $a_3=-0.9$ and the dashed line is for $a_3=-1.1$. 
The $K^-$ energy decreases with density due to the $s$-wave 
kaon-baryon interaction in $\Pi_K^s$ [Eq.~(\ref{eq:selfs})].  
However, 
due to the substantial decrease of the charge chemical potential 
$\mu$ with density in the presence of hyperons, the onset condition  
for the $s$-wave $K^-$ condensation, 
$\omega_{\rm min}(K^-)=\mu$, is met at a much larger density (filled circles) 
than that in the conventional neutron-star matter which consists of only 
the nucleons $n$, $p$ and $e^-$. On the other 
hand, the critical density of the $p$-wave condensation is indicated 
by the arrows in Fig.~\ref{fig:swave} (the solid arrow for $a_3=-0.9$ 
and the dashed arrow for $a_3=-1.1$). One finds that the $p$-wave 
condensation precedes the $s$-wave $K^-$ condensation. 

\begin{figure}[t]
\noindent\begin{minipage}[l]{0.48\textwidth}
\centerline{
\epsfxsize=\textwidth\epsffile{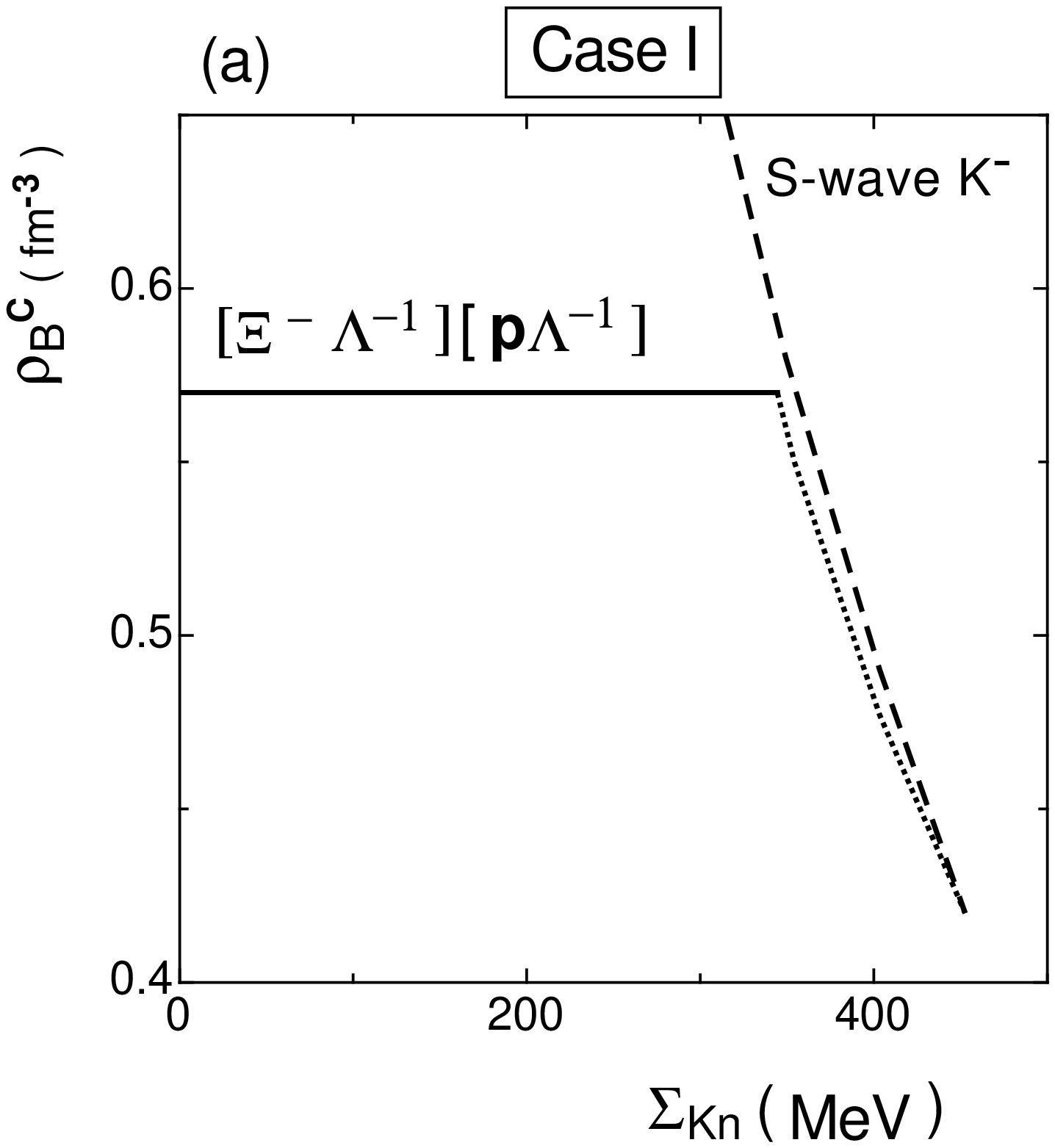}}
\end{minipage}~
\begin{minipage}[r]{0.50\textwidth}
\centerline{
\epsfxsize=\textwidth\epsffile{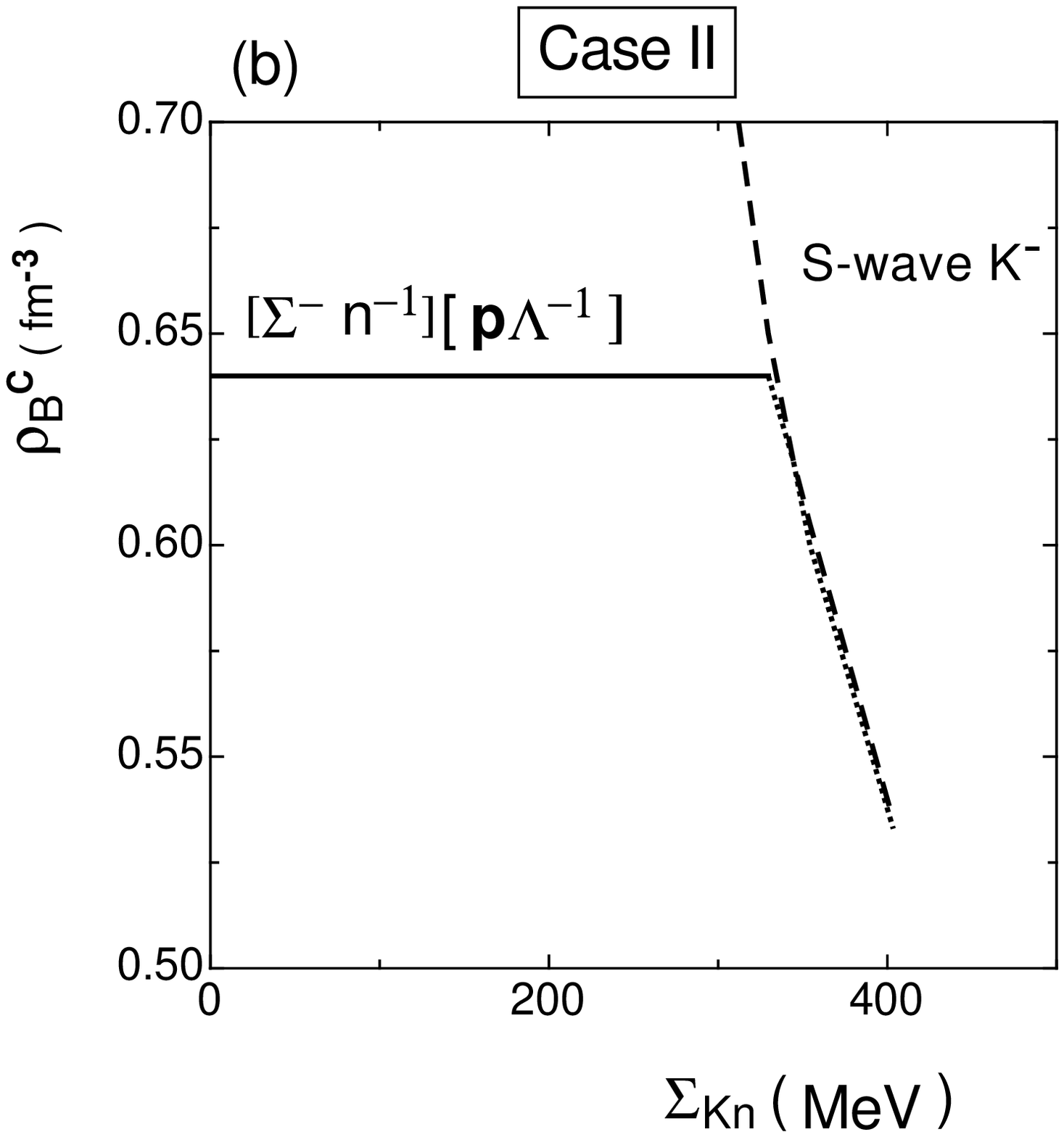}}
\end{minipage}
\caption{(a) The critical density of the $p$-wave kaon 
condensation (the solid and dotted lines) 
as a function of the kaon-neutron sigma term $\Sigma_{Kn}$ in Case I. 
For comparison, the critical density for the $s$-wave $K^-$ 
condensation is shown by the dashed line. 
(b) The same as (a), but for Case II. See the text for the details. }
\label{fig:rhoc}
\end{figure}

In Fig.~\ref{fig:rhoc}, we summarize the dependence of the critical 
density of the $p$-wave kaon condensation $\rho_{\rm B}^{\rm C}(p)$ 
(solid and dotted lines) on the kaon-neutron sigma term $\Sigma_{Kn}$ 
[$\equiv -(a_2+2a_3)(m_u+m_s)$] .  For comparison, the critical 
density of the $s$-wave $K^-$ condensation $\rho_{\rm B}^{\rm C}(s)$ 
(the dashed line) is also shown. (a) is for Case I, and (b) is for 
Case II. For the strong $s$-wave scalar attraction such that 
$\Sigma_{Kn}\gtrsim$ 340 MeV ($a_3\lesssim -0.98$), there are level 
crossings between the $K^-$ and the particle-hole branches. As a 
result, the critical density $\rho_{\rm B}^{\rm C}(p)$ is sensitive 
to the magnitude of $|a_3|$. $\rho_{\rm B}^{\rm C}(p)$ is slightly 
smaller than $\rho_{\rm B}^{\rm C}(s)$ in both Cases I and II, but the 
difference is small, as shown by the dotted lines and dashed lines.  
For the weaker $s$-wave 
attraction such that $\Sigma_{Kn}\lesssim$ 340 MeV ($a_3\gtrsim 
-0.98$), the critical density of the 
$p$-wave condensation $\rho_{\rm B}^{\rm C}(p)$ depends little on the 
magnitude of $a_3$ (the solid lines). This is because the 
$p\Lambda^{-1}$, $\Xi^-\Lambda^{-1}$. and $\Sigma^- n^{-1}$ branches 
hardly depend on the magnitude of $a_3$ as long as no level crossing 
with the $K^-$ branch occurs before the onset of instability. The 
critical density $\rho_{\rm B}^{\rm C}(p)$ changes little over the 
range from $\Sigma_{Kn}$=0 ($a_3=-0.28$) to $\Sigma_{Kn}\sim $ 340 
MeV ($a_3\sim -0.98$) in both Cases I and II. 
On the other hand, the critical density for the $s$-wave $K^-$ 
condensation $\rho_{\rm B}^{\rm C}(s)$ becomes large as $\Sigma_{Kn}$ 
(or $|a_3|$) becomes small, due to the reduced contribution from the 
$s$-wave scalar attraction to the $K^-$ excitation energy 
$\omega(K^-)$. In conclusion, the critical density of the $p$-wave 
condensation is always smaller than that of the $s$-wave $K^-$ 
condensation as long as $|a_3|$ is not too large, and the difference 
between these critical densities gets remarkable with the decrease 
in the magnitude of the $s$-wave scalar attraction. 

\subsection{Outline of the condensed phase}
\label{subsec:outline}

We address qualitative features of the $p$-wave kaon condensation discussed in this paper. 
The details will be given elsewhere\cite{m01}. 

At a critical density $\rho_{\rm B}^{\rm C}$, pairs  of the kaonic modes, 
[$p\Lambda^{-1}$]-[$\Xi^-\Lambda^{-1}$] in Case I or [$p\Lambda^{-1}$]-[$\Sigma^-n^{-1}$] 
in Case II, are spontaneously created via the reaction, 
$\Lambda \Lambda\rightarrow \Xi^- p$ in Case I or $\Lambda n\rightarrow \Sigma^- p$
in Case II. Thus, formation of a condensate in hyperonic matter proceeds 
essentially through the strong reactions, which 
should be compared with the $s$-wave $K^-$ condensation: 
In the latter case, a condensate is formed through the weak reactions 
such as $nn\rightarrow npK^-$, $e^-\rightarrow K^-\nu_e$. 
In the $p$-wave condensed phase, baryonic system consists of the Fermi seas 
of the quasiparticles $\tilde p$, $\tilde\Lambda$, $\tilde\Xi^-$, $\tilde n$, 
$\tilde \Sigma^-$ (see Sec.~\ref{subsec:energy}).  The absolute value of total 
negative strangeness of the system increases as the baryon number density 
increases by virtue of the weak reactions. The relevant strangeness-changing
weak processes are given by $\tilde Y \langle K^-\rangle\rightarrow  \tilde Y e^-\bar\nu_e$, 
$\tilde Y e^-\rightarrow \tilde Y \langle K^-\rangle \nu_e$ (for $\tilde Y$=$\tilde p$, 
$\tilde\Lambda$, $\tilde\Xi^-$, $\tilde n$, $\tilde \Sigma^-$), where the classical 
kaon field $\langle K^-\rangle$ given by Eq.~(\ref{eq:classical}) supplies the system 
with the energy and momentum to satisfy the kinematical condition for these reactions. 
These weak processes may also be relevant to enhanced cooling of neutron stars 
via neutrino emissions just like a case of pion condensed phase\cite{m78,bc79,kmttt93}. 

The EOS of the $p$-wave kaon-condensed phase would be further softened 
as compared with that of noncondensed hyperonic matter owing to the 
$p$-wave kaon-baryon attractive interaction in addition to the $s$-wave one. 
The significant softening will make phase transition of a first order 
after Maxwell's construction, leading to a drastic change of the internal structure 
of neutron stars. In particular, the first-order phase transition may imply 
a mixed phase where droplets of a kaon condensate are immersed 
in the normal phase\cite{gs98}. 
The first-order phase transition has also important effects on the dynamical 
evolution of newly-born neutron stars accompanying a delayed collapse, 
which has already been discussed by several authors in case of the $s$-wave kaon 
condensation\cite{ty98,p00}. 
\section{Summary and concluding remarks}
\label{sec:summary}

We have discussed in-medium properties of kaonic modes in hyperonic 
matter by taking into account the $p$-wave kaon-baryon interaction as 
well as the $s$-wave one on the basis of chiral symmetry. 
Nonrelativistic effective baryon-baryon interactions,  which are 
parametrized by the use of the recent hypernuclear experimental data, 
have been used. It has been shown that a collective $p\Lambda^{-1}$ 
mode with the $K^+$ quantum number appears over the densities where 
the $\Lambda$ is more abundant than the proton. The system becomes 
unstable with respect to a creation of 
[$\Xi^- \Lambda^{-1}$] and [$p\Lambda^{-1}$] pair 
or [$\Sigma^-n^{-1}$] and  [$p\Lambda^{-1}$] pair ($p$-wave kaon 
condensation), which stems from the $p$-wave kaon-baryon interaction. 
The onset density of this instability is lower than that of the 
$s$-wave $K^-$ condensation for a standard value of the parameter 
$a_3$ simulating the magnitude of the $s$-wave kaon-baryon scalar 
interaction, and it hardly depends on the value of $a_3$ as long as 
$|a_3|$ is not too large. 

The possibility of the $p$-wave kaon condensation depends on 
composition of baryons in hyperonic matter. In particular,  
large mixing of $\Lambda $ as compared with that of the proton is 
needed for the appearance of the $p\Lambda^{-1}$ mode. 
The details about the onset densities of hyperons and their fractions 
at high densities differ between specific models for the baryonic  
potentials. One of the important ingredients which control matter 
composition is three-body forces for baryons\cite{bbs98,y00}. It has 
been shown that phenomenological inclusion of  three-body forces for 
only nucleons makes hyperon-mixing  favorable\cite{bbs98}. However, 
it has been pointed out that inclusion of three-body forces for 
hyperons on the same footing as the nucleons may considerably change 
the results on the matter composition and the resultant EOS of the 
hyperonic matter\cite{y00}. 
Thus one has to be careful for the parametrization of the effective 
baryon-baryon interactions used in this paper, keeping consistency 
with these other model calculations. 

Our model used for the $p$-wave kaon-baryon interaction is based on 
the leading order expansion in the chiral perturbation theory. 
Higher order terms in chiral expansion which are relevant to the 
$p$-wave meson-baryon scatterings have been estimated with reference 
to the experimental results such as pion and photon-induced 
reactions\cite{rkww00} or elastic and inelastic $K^- p$ 
scatterings\cite{lk00}.  It needs more consideration whether these 
higher order terms are quantitatively important to kaon dynamics in 
highly dense matter. 

It has to be elucidated whether the instability of the system with 
respect to the $p$-wave condensation leads to a fully condensed phase 
beyond  the critical density. In this context, the EOS of the 
$p$-wave condensed phase and the characteristic features of the 
system have to be examined\cite{m01}. Mixing of hyperons only already 
leads to appreciable softening of the 
EOS\cite{g85,h98,bbs98,v00,bg97,blc99,y00,phz99,h00,kj95,p99}. Hence, 
further development of kaon condensates in hyperonic matter would 
make the EOS too soft to obtain the observed neutron star masses 
$\sim 1.4 M_\odot$\cite{tw89,tc99} or even much larger masses$\sim 
2.0 M_\odot$ if the recent analyses from the observations of the 
quasi-periodic oscillations (QPO)\cite{mlp98} are 
confirmed\cite{v00}.  Relativistic 
effects  may help weaken the softness of the EOS, since it has been 
shown that the energy gain of kaon condensation coming from the 
$s$-wave scalar attraction is suppressed by relativistic 
effects\cite{mfmt94,fmmt96}: As a kaon condensate develops, the 
effective nucleon mass $M^\ast$ decreases due to the scalar 
attraction, which leads to suppression of the scalar density, 
$\displaystyle \rho_s=\int \frac{d^3 p}{(2\pi)^3}M^\ast/\sqrt{{\bf 
p}^2+M^{\ast 2}}$. Thus the energy gain of the condensed phase from 
the scalar attraction ($\propto\rho_s\Sigma_{KN}$ with the $KN$ sigma 
term $\Sigma_{KN}$) and the growth of a condensate are  suppressed. 

In addition, in view of making the EOS consistent with observations,  
some realistic effects which reduce the $p$-wave 
kaon-baryon attraction should be taken into account: (1) vertex 
renormalization at the $p$-wave 
kaon-baryon vertices in terms of form factors. 
(2) short-range correlations between baryons. For the $p$-wave part, 
the off-diagonal matrix elements in the baryonic part of the 
effective Hamiltonian [Eq.~(\ref{eq:bhm})] are to be added by the 
particle-hole densities, the strengths of which are related with the 
Landau-Migdal parameters in the relevant channel in the same way as pion 
condensation\cite{ab74,m78,bc79,ew88,kmttt93}. 

\section{Acknowledgements}
\label{sec:ack}
The author would like to thank T. Tatsumi for useful discussions and 
comments. The author is also  grateful to T. Takatsuka, S. Nishizaki, 
M. Yasuhira, D. N. Voskresensky, J. Schaffner-Bielich and P. K. Sahu 
for discussions and interest in this work. 
This work is supported in part by the Japanese Grant-in-Aid for 
Scientific Research Fund (C) of the Ministry of Education, Science, 
Sports, and Culture (No. 12640289). Numerical calculations were 
carried out on the DEC Alpha Server 4100 System, Chiba Institute of 
Technology. 

\appendix
\section{Potential energy density in hyperonic matter}
\label{sec:app}

We show the expression for the potential energy density 
${\cal E}_{\rm pot}$ based on the nonrelativistic baryon-baryon 
interactions by Balberg and Gal (Eq.~(6) in \cite{bg97}). Since only 
$p$, $\Lambda$, $\Xi^-$, $n$, and $\Sigma^-$ are incorporated for the 
baryons in this paper, the other terms relevant to $\Sigma^0$, 
$\Sigma^+$, and $\Xi^0$ are omitted.  
\begin{eqnarray}
{\cal E}_{\rm pot}&=&\frac{1}{2}\Big\lbrack a_{\rm NN}(\rho_{\rm 
p}+\rho_{\rm n})^2+b_{\rm NN}(\rho_{\rm p}-\rho_{\rm n})^2
+c_{\rm NN}(\rho_{\rm p}+\rho_{\rm n})^{\delta+1} \Big\rbrack\cr
&+& a_{\rm \Lambda N}(\rho_{\rm p}+\rho_{\rm n}){\rho_\Lambda}+c_{\rm 
\Lambda N}\Bigg\lbrack\frac{(\rho_{\rm p}
+\rho_{\rm n})^{\gamma+1}}{\rho_{\rm p}+\rho_{\rm 
n}+{\rho_\Lambda}}{\rho_\Lambda}
+\frac{{\rho_\Lambda}^{\gamma+1}}{\rho_{\rm p}
+\rho_{\rm n}+{\rho_\Lambda}}(\rho_{\rm p}
+\rho_{\rm n})\Bigg\rbrack \cr
&+&\frac{1}{2}\Big\lbrack a_{YY}{\rho_\Lambda}^2
+c_{\rm YY}{\rho_\Lambda}^{\gamma+1}+(a_{\rm YY}+
b_{\Xi\Xi}){\rho_{\Xi^-}}^2+c_{\rm 
YY}{\rho_{\Xi^-}}^{\gamma+1}\Big\rbrack \cr 
&+&a_{\rm \Xi N}(\rho_{\rm p}+\rho_{\rm n}){\rho_{\Xi^-}}
+b_{\rm \Xi N}(\rho_{\rm n}-\rho_{\rm p}){\rho_{\Xi^-}} \cr
&+&c_{\rm \Xi N}\Bigg\lbrack\frac{(\rho_{\rm p}
+\rho_{\rm n})^{\gamma+1}}{\rho_{\rm p}
+\rho_{\rm n}+\rho_{\Xi^-}}\rho_{\Xi^-}
+\frac{{\rho_{\Xi^-}}^{\gamma+1}}{\rho_{\rm p}
+\rho_{\rm n}+{\rho_{\Xi^-}}}(\rho_{\rm p}
+\rho_{\rm n})\Bigg\rbrack \cr
&+&a_{\rm YY}{\rho_{\Xi^-}}{\rho_\Lambda}
+c_{\rm YY}\Bigg\lbrack 
\frac{\rho_\Lambda^{\gamma+1}}{\rho_{\Xi^-}+\rho_\Lambda}\rho_{\Xi^-}
+\frac{{\rho_{\Xi^-}}^{\gamma+1}}{{\rho_{\Xi^-}}+{\rho_\Lambda}}\rho_\Lambda\Bigg\rbrack 
\cr
&+&a_{\rm \Sigma N}(\rho_{\rm p}+\rho_{\rm 
n}){\rho_{\Sigma^-}}+b_{\rm \Sigma N}(\rho_{\rm n}-\rho_{\rm 
p}){\rho_{\Sigma^-}} \cr 
&+& c_{\rm \Sigma N}\Bigg\lbrack\frac{(\rho_{\rm p}
+\rho_{\rm n})^{\gamma+1}}{\rho_{\rm p}+\rho_{\rm 
n}+{\rho_{\Sigma^-}}}{\rho_{\Sigma^-}}
+\frac{{\rho_{\Sigma^-}}^{\gamma+1}}{\rho_{\rm p}
+\rho_{\rm n}+\rho_{\Sigma^-}}(\rho_{\rm p}
+\rho_{\rm n})\Bigg\rbrack \cr
&+&a_{\rm YY}{\rho_{\Sigma^-}}{\rho_\Lambda}+c_{\rm YY}\Bigg\lbrack 
\frac{{\rho_{\Sigma^-}}^{\gamma+1}}{{\rho_{\Sigma^-}}+{\rho_\Lambda}}{\rho_\Lambda}+\frac{{\rho_\Lambda}^{\gamma+1}}{{\rho_{\Sigma^-}}+{\rho_\Lambda}}{\rho_{\Sigma^-}}\Bigg\rbrack 
\cr
&+& a_{\rm 
YY}{\rho_{\Sigma^-}}{\rho_{\Xi^-}}+b_{\Sigma\Xi}{\rho_{\Xi^-}}{\rho_{\Sigma^-}}+c_{\rm 
YY}\Bigg\lbrack 
\frac{{\rho_{\Xi^-}}^{\gamma+1}}{{\rho_{\Xi^-}}+{\rho_{\Sigma^-}}}{\rho_{\Sigma^-}}+\frac{{\rho_{\Sigma^-}}^{\gamma+1}}{{\rho_{\Xi^-}}+{\rho_{\Sigma^-}}}{\rho_{\Xi^-}}\Bigg\rbrack 
\cr
&+&\frac{1}{2}\Big\lbrack (a_{\rm YY}
+b_{\Sigma\Sigma}){\rho_{\Sigma^-}}^2
+c_{\rm YY}{\rho_{\Sigma^-}}^{\gamma+1}\Big\rbrack 
\label{eq:epot}
\end{eqnarray}
The parameters relevant to the $NN$ and $YN$ parts in 
Eq.~(\ref{eq:epot}) have been refitted in reference to the recent 
empirical data on the nuclear and hypernuclear properties. Numerical 
values of the parameters are listed in Tables \ref{tab:para} and 
\ref{tab:paras}.


\begin{table}
\caption{Parameters in the potential energy density. ($^{\rm 
a}$MeV$\cdot$fm$^3$, \  $^{\rm b}$MeV$\cdot$fm$^{3\gamma}$)
}
\label{tab:para}
\begin{tabular}{cc|cc|cc}
parameter & & parameter & & parameter & \\ \hline
 $\gamma$ & 5/3 &  ${a_{\Lambda N}}^{\rm a}$ & $-$387.0 & 
${a_{YY}}^{\rm a}$ & $-$552.6 \\
$\delta$ & 5/3 & ${c_{\Lambda N}}^{\rm b}$ & 738.8 & ${c_{YY}}^{\rm 
b}$ & 1055.4   \\
${a_{NN}}^{\rm a}$ & $-$859.5 & ${a_{\Xi N}}^{\rm a}$ & $-$228.6 &  
${b_{\Xi\Xi}}^{\rm a}$ & 0  \\
${b_{NN}}^{\rm a}$ & 212.8 & ${b_{\Xi N}}^{\rm a}$ & 0 & 
${b_{\Sigma\Xi}}^{\rm a}$ & 0  \\
${c_{NN}}^{\rm b}$ & 1300.8 & ${c_{\Xi N}}^{\rm b}$ & 436.5 & 
${b_{\Sigma\Sigma}}^{\rm a}$ & 428.4 \\
\end{tabular}
\end{table}

\begin{table}
\caption{Parameters for the $\Sigma^- N$ part in the potential energy 
density. ( $^{\rm a}$MeV$\cdot$fm$^3$, \ 
$^{\rm b}$MeV$\cdot$fm$^{3\gamma}$)}
\label{tab:paras}
\begin{tabular}{ccc}
 & Case I & Case II \\ \hline
 ${a_{\Sigma N}}^{\rm a}$ & $-$70.9 & $-$387.0 \\
 ${b_{\Sigma N}}^{\rm a}$ & 251.3 & 214.2 \\
 ${c_{\Sigma N}}^{\rm b}$ & 738.8 & 738.8 \\
\end{tabular}
\end{table}


\begin{thebibliography}{9}
{\footnotesize
\bibitem{kn86}D.~B.~Kaplan and A.~E.~Nelson, Phys.~Lett.~{\bf B175}
 (1986) 57; {\bf 179} (1986) 409(E). 
 \bibitem{lbm95} C.~-H.~Lee, G.~E.~Brown, D.~-P.~Min and M.~Rho, 
 Nucl.~Phys.~{\bf A 585} (1995) 401. 
 \bibitem{t95}
For a review, T. Tatsumi, Prog. Theor. Phys. Suppl. {\bf 120}
(1995) 111. 
\bibitem{fmmt96}
H.~Fujii, T.~Maruyama, T.~Muto and T.~Tatsumi, 
Nucl. Phys. {\bf A597} (1996) 645 .
\bibitem{l96}  C.~-H.~Lee, Phys.~Rep.~{\bf 275} (1996)197 . 
\bibitem{pbpelk97} M.~Prakash, I.~Bombaci, M.~Prakash, P.~J.~Ellis,
J.~M.~Lattimer, R.~Knorren, Phys. Rep. {\bf 280} (1997) 1. 
\bibitem{tpl94} V.~Thorsson, M.~Prakash and J.~M.~Lattimer, 
Nucl.~Phys.~{\bf
A 572} (1994), 693 ; {\it ibid}~{\bf A 574} (1994), 851 (E). 
\bibitem{mfmt94} T.~Maruyama, H.~Fujii, T.~Muto and T.~Tatsumi, 
Phys.~Lett.~{\bf B 337} (1994) 19.
\bibitem{gs98} N.~K.~Glendenning and J.~Schaffner-Bielich, 
Phys.~Rev.~Lett.~{\bf 81} (1998) 4564 ; 
Phys.~Rev.~{\bf C60} (1999) 025803 . 
\bibitem{ty98} T.~Tatsumi and M.~Yasuhira, Nucl.~Phys.~{\bf A 653} 
(1999) 133; {\it ibid} {\bf A 670} (2000) 218; 
M.~Yasuhira and T.~Tatsumi, Nucl.~Phys.~{\bf A 663} (2000) 881c. 
\bibitem{p00} J.~A.~Pons, S.~Reddy, P.~J.~Ellis, M.~Prakash and 
J.~M.~Lattimer, Phys.~Rev.~{\bf C 62} (2000) 035803. 
\bibitem{bkpp88} G.~E.~Brown, K.~Kubodera, D.~Page and 
P.~Pizzecherro, 
Phys.~Rev.~{\bf D37} (1988) 2042 . 
\bibitem{t88} T.~Tatsumi, Prog.~Theor.~Phys.~{\bf 80} (1988) 22 . 
\bibitem{pb90} D.~Page and E.~Baron, 
Astrophys.~J.~{\bf 254} (1990) L17. 
\bibitem{fmtt94} H.~Fujii, T.~Muto, T.~Tatsumi and R.~Tamagaki, 
Nucl.~Phys.~{\bf A571},758 (1994); Phys.~Rev.~{\bf C50} (1994)  3140. 
\bibitem{wrw97} T.~Waas, M.~Rho and W.~Weise, Nucl.~Phys.~{\bf A 617} 
(1997) 449.
\bibitem{chp00} J.~Carlson, H.~Heiselberg and V.~R.~Pandharipande, 
Phys.~Rev.~{\bf C 63} (2000) 017603. 
\bibitem{k94} V.~Koch, Phys.~Lett.~{\bf B337} (1994) 7 .
\bibitem{wkw96} T.~Waas,N.~Kaiser and W.~Weise, 
Phys.~Lett.~{\bf B365} (1996)12 ; {\bf B379} (1996) 34 . \\ 
T.~Waas and W.~Weise, Nucl.Phys.~{\bf A625} (1997) 287. 
\bibitem{l98} M.~Lutz, Phys.~Lett.~{\bf B426} (1998) 12 . 
\bibitem{ro00} A.~Ramos and E.~Oset, Nucl.~Phys.~{\bf A 671} (2000) 
481. 
\bibitem{fgb94} E.~Friedman, A.~Gal and C.~J.~Batty, Nucl.~Phys.~{\bf 
A 579} (1994) 518; \\
C.~J.~Batty, E.~Friedman and A.~Gal,  Phys.~Rep.~{\bf 287} (1997) 
385. 
\bibitem{fgm99} E.~Friedman, A.~Gal, J.~Mares and A.~Cieply, 
Phys.~Rev.~{\bf C 60} (1999) 024314.
\bibitem{ho00} S.~Hirenzaki, Y.~Okumura, H.~Toki, E.~Oset and 
A.~Ramos, Phys.~Rev.~{\bf C 61} (2000) 055205. 
\bibitem{brn00} A.~Baca, C.~Garcia-Recio and J.~Nieves, 
Nucl.~Phys.~{\bf A 673} (2000) 335. 
\bibitem{llb97}G.~Q.~Li, C.-~H.~Lee and G.~E.~Brown, 
Phys.~Rev.~Lett.~{\bf 79} (1997) 5214 ; 
Nucl.~Phys.~{\bf A625} (1997) 372. 
\bibitem{kl96} C.~M.~Ko and G.~Q.~Li, J.~Phys.~{\bf G 22} (1996) 
1673.
\bibitem{slk99} G.~Song, B.-~A.~Li and C.~M.~Ko, 
Nucl.~Phys.~{\bf A646} (1999) 481. 
\bibitem{smb97} J.~Schaffner-Bielich, I.~N.~Mishustin and 
J.~Bondorf, Nucl.~Phys.~{\bf A625} (1997) 325 . 
\bibitem{cb99} W.~Cassing and E.~L.~Bratkowskaya, 
Phys.~Rep.~{\bf 308} (1999) 65 . 
\bibitem{sc98} A.~Sibirtsev and W.~Cassing, 
Nucl.~Phys.~{\bf A 641} (1998) 476; nucl-th/9909024. 
\bibitem{ske00} J.~Schaffner-Bielich, V.~Koch and M.~Effenberger, 
Nucl.~Phys.~{\bf A 669} (2000) 153.
\bibitem{trp00} L.~Tol{\'o}s, A.~Ramos, A.~Polls and T.~T.~S.~Kuo, 
nucl-th/0007042.
\bibitem{b97} R.~Barth et al., Phys.~Rev.~Lett.{\bf 78} (1997) 4007. 
\bibitem{l99} F.~Laue et al., Phys.~Rev.~Lett.{\bf 82} (1999) 1640. 
\bibitem{ss99} For a review, 
P.~Senger and H.~Str{\"o}bele, J.~Phys.~G{\bf 25} (1999) R59. 
\bibitem{k99} T.~Kishimoto, Phys.~Rev.~Lett.~{\bf 83} (1999) 4701. 
\bibitem{a00} Y.~Akaishi, 
Proceedings of the VII International 
Conference on Hypernuclear and Strange Particle Physics, Torino, 
Italy, October 23-27,2000, to be published in Nucl.~Phys.~{\bf A}; 
M.~Iwasaki, {\it ibid}. 
\bibitem{fg00} E.~Friedman and A.~Gal, Nucl.~Phys.~{\bf A 663} (2000) 557. 
\bibitem{c59} A.~G.~W.~Cameron, Astrophys.~J.~{\bf 130} (1959) 884. 
\bibitem{as60} V.~A.~Ambartsumyan and G.~S.~Saakyan, Sov.~Astron.~{\bf 4} (1960) 187. 
\bibitem{tc66} S.~Tsuruta and A.~G.~W.~Cameron, Can.~J.~Phys.~{\bf 
44} (1966) 1895.
\bibitem{lr70} W.~D.~Langer and L.~Rosen, 
Astrophys.~Space Sci.~{\bf 6} (1970) 217.
\bibitem{p71} V.~R.~Pandharipande, 
Nucl.~Phys.~{\bf A 178} (1971) 123.
\bibitem{g85} N.~K.~Glendenning, 
Astrophys.~J.~{\bf 293} (1985) 470 ; N.~K.~Glendenning and S.~A.~Moszkowski, 
Phys.~Rev.~Lett.~{\bf 67} (1991) 2414. 
\bibitem{ekp95} P.~J.~Ellis, R.~Knorren, and M.~Prakash, 
Phys.~Lett.~{\bf B349} (1995) 11 . \\
R.~Knorren, M.~Prakash and P.~J.~Ellis, 
Phys.~Rev.~{\bf C52} (1995) 3470. 
\bibitem{sm96} J.~Schaffner and I.~N.~Mishustin, 
Phys.~Rev.~{\bf C53} (1996) 1416. 
\bibitem{s00} P.~K.~Sahu, Phys.~Rev.~{\bf C 62} (2000) 045801. 
\bibitem{h98} H.~Huber, F.~Weber, M.~K.~Weigel and Ch.~Schaab, 
Int.~J.~Mod.~Phys.~{\bf E 7} (1998) 301. 
\bibitem{bbs98} M.~Baldo, G.~F.~Burgio, and H.~-J.~Schulze, 
Phys.~Rev.~{\bf C 58} (1998) 3688 ; {\it ibid} {\bf C  61} (2000) 
055801.
\bibitem{v00} I.~Vida$\tilde{\rm n}$a, A.~Polls, A.~Ramos, 
M.~Hjorth-Jensen and V.~G.~J.~Stoks, Phys.~Rev.~{\bf C 61} (2000) 
025802; 
I.~Vida$\tilde {\rm n}$a, A.~Polls, A.~Ramos, L.~Engvik and 
M.~Hjorth-Jensen, Phys.~Rev.~{\bf C 62} (2000) 035801. 
\bibitem{bg97} S.~Balberg and A.~Gal, Nucl.~Phys.~{\bf A625} (1997) 
435. 
\bibitem{blc99} S.~Balberg, I.~Lichtenstadt and G.~B.~Cook, 
Astrophys.~J.~Suppl. {\bf 121} (1999) 515. 
 \bibitem{y00} S.~Nishizaki, Y.~Yamamoto and T.~Takatsuka,  
 Prog.~Theor.~Phys.~{\bf 105} (2001) 607; {\it ibid} (2001), submitted. 
\bibitem{phz99} S.~Pal, M.~Hanauske, I.~Zakout, H.~St{\"o}cker and 
W.~Greiner, Phys.~Rev.~{\bf C 60} (1999) 015802. 
\bibitem{h00} M.~Hanauske, D.~Zschiesche, S.~Pal, S.~Schramm, 
H.~St{\"o}cker and W.~Greiner, Astrophys.~J.~{\bf 537} (2000) 958. 
\bibitem{kj95} W.~Keil and H.-Th.~Janka, Astron.~Astrophys.~{\bf 
296} (1995) 145 .
\bibitem{p99} J.~A.~Pons,  et al, 
Astrophys.~J.~{\bf 513} (1999) 780.  
\bibitem{ppl92} M.~Prakash, M.~Prakash, J.~M.~Lattimer, and 
C.~J.~Pethick, Astrophys.~J.~{\bf 390} (1992) L77. 
\bibitem{blr98} G.~E.~Brown, C.-~H.~Lee and R.~Rapp, 
Nucl.~Phys. {\bf A639} (1998) 455c. 
\bibitem{m93} T.~Muto, Prog.~Theor.~Phys.~{\bf 89} (1993) 415.
\bibitem{kvk95} E.~E.~Kolomeitsev, D.~N.~Voskresensky and 
B.~K{\"a}mpfer, Nucl.~Phys.~{\bf A588} (1995) 889. 
\bibitem{m00} T.~Muto, Proceedings of the VII International 
Conference on Hypernuclear and Strange Particle Physics, Torino, 
Italy, October 23-27,2000, to be published in Nucl.~Phys.~{\bf A}. 
\bibitem{ynm93} H.~Yabu, S.~Nakamura, F.~Myhrer and K.~Kubodera, 
Phys.~Lett.~{\bf B 315} (1993) 17. 
\bibitem{ljm94} C.~H.~Lee, H.~Jung, D.~-P.~Min and M.~Rho, 
Phys.~Lett.~{\bf B 326} (1994) 14. 
\bibitem{ymk94} H.~Yabu, F.~Myhrer and K.~Kubodera, Phys.~Rev.~{\bf D 
50} (1994) 3549. 
\bibitem{tw95} V.~Thorsson and A.~Wirzba, Nucl.~Phys.~{\bf A 589} 
(1995) 633.
 \bibitem{m01} T.~Muto, unpublished. 
\bibitem{rkww00} J.~C.~Ramon, N.~Kaiser, S.~Wetzel and W.~Weise, 
Nucl.~Phys.~{\bf A 672} (2000) 249. 
\bibitem{lk00} M.~F.~M.~Lutz and E.~E.~Kolomeitsev, nucl-th/0004021. 
\bibitem{ab74} C.~-K.~Au and G.~Baym, Nucl.~Phys.~{\bf A 236} (1974) 
500. 
\bibitem{mo88} R.~Mittet and E.~{\O}stgaard, Phys.~Rev.~{\bf C 37} 
(1988) 1711. 
\bibitem{b81} J.~Boguta, Phys.~Lett.~{\bf B 106} (1981) 255. 
\bibitem{lpph91} J.~M.~Lattimer, C.~J.~Pethick, M.~Prakash and 
P.~Haensel, Phys.~Rev.~Lett.~{\bf 66} (1991) 2701. 
\bibitem{mdg98} D.~J.~Millener, C.~B.~Dover and A.~Gal, 
Phys.~Rev.~{\bf C 58} (1998) 22700. 
\bibitem{f98} T.~Fukuda et al., Phys.~Rev.~{\bf C58} (1998) 1306. 
\bibitem{k00} P.~Khaustov et al., Phys.~Rev.~{\bf C 61} (2000) 
054603. 
\bibitem{d99} J.~Dabrowski, Phys.~Rev.~{\bf C 60} (1999) 025205. 
\bibitem{b99} S.~Bart et al., Phys.~Rev.~Lett.~{\bf 83} (1999) 5238. 
\bibitem{mfgj95} J.~Mares, E.~Friedman, A.~Gal and B.~K.~Jennings, 
Nucl.~Phys.~{\bf A 594} (1995) 311. 
\bibitem{kf00} M.~Kohno, Y.~Fujiwara, T.~Fujita, C.~Nakamoto and 
Y.~Suzuki, Nucl.~Phys.~{\bf A 674} (2000) 229. 
\bibitem{m78} A.~B.~Migdal, Rev.~Mod.~Phys.~{\bf 50} (1978) 107. \\
 A.~B.~Migdal, E.~E.~Saperstein, M.~A.~Troitsky and 
D.~N.~Voskresensky, Phys.~Rep.~{\bf 192} (1990) 179. 
\bibitem{bc79} G.~Baym and D.~K.~Campbell, in {\it Meson and Nuclei}, 
ed. M.~Rho and D.~H.~Wilkinson, (North Holland, Amsterdam, 1979), 
Vol.~III, p.~1031.
\bibitem{ew88} T.~Ericson and W.~Weise, ``Pions and Nuclei", 
(Clarendon Press, Oxford, 1988).
\bibitem{kmttt93} T.~Kunihiro, T.~Muto, T.~Takatsuka, R.~Tamagaki and 
T.~Tatsumi, \\
Prog.~Theor.~Phys.~Suppl. {\bf 112} (1993). 
\bibitem{tw89} J.~H.~Taylor and J.~M.~Weisberg, Astrophys.~J.~{\bf 
345} (1989) 434. 
\bibitem{tc99} S.~E.~Thorsett and D.~Chakrabarty, 
Astrophys.~J.~{\bf 512} (1999) 288. 
\bibitem{mlp98} M.~C.~Miller, F.~K.~Lamb and D.~Psaltis, 
Astrophys.~J.~{\bf 508} (1998) 791. 
}
\end{thebibliography}
\end{document}